\documentclass[twocolumn,showpacs,amsmath,amssymb,pra,aps]{revtex4}

\usepackage{amsmath}
\usepackage{graphicx}
\usepackage{bm}

\begin{document}

\title{%
Electromagnetic-field quantization and spontaneous decay in
left-handed media
}
\author{Ho Trung Dung}
\altaffiliation[Also at]{
Institute of Physics, National Center
for Sciences and Technology, 1 Mac Dinh Chi Street,
District 1, Ho Chi Minh city, Vietnam.}

\author{Stefan Yoshi Buhmann}

\author{Ludwig Kn\"{o}ll}

\author{Dirk-Gunnar Welsch}

\affiliation{Theoretisch-Physikalisches Institut,
Friedrich-Schiller-Universit\"at Jena,
Max-Wien-Platz 1, D-07743 Jena, Germany}

\author{Stefan Scheel}
\affiliation{Quantum Optics and Laser Science, Blackett Laboratory,
Imperial College London, Prince Consort Road, London SW7 2BW, United Kingdom}

\date{June 4, 2003}

\begin{abstract}
We present a quantization scheme for the electromagnetic field
interacting with atomic systems in the presence of
dispersing and absorbing magnetodielectric media, including
left-handed material
having negative real part of the refractive index.
The theory is applied to the spontaneous decay of a two-level atom 
at the center of a spherical free-space cavity surrounded
by magnetodielectric matter of overlapping band-gap zones.
Results for both big and small cavities are presented,
and the problem of local-field corrections within
the real-cavity model is addressed.
\end{abstract}

\maketitle


\section{Introduction}
\label{introduction}

The problem of propagation of electromagnetic waves in
materials having, in a certain frequency range,
simultaneously negative permittivity and permeability thus leading
to a negative refractive index was first studied by
Veselago \cite{Veselago68}.
Since in such materials the electric field, the
magnetic field, and wave vector of a plane
wave form a left-handed system, so that the direction of the
Poynting vector and the wave vector have opposite directions,
they are also called left-handed materials (LHMs).
Other unusual properties are a reverse Doppler shift,
reverse Cerenkov radiation, negative refraction, and
reverse light pressure.
Since LHMs do not exist naturally, they have remained a
merely academic curiosity until recent reports on their fabrication
\cite{Smith00a,Shelby01sci,Marques02,Grbic02,Parazzoli03}.
The metamaterials considered
there
consist of periodic arrays of metallic thin wires to attain negative
permittivity, interspersed with split-ring resonators
to attain negative permeability. Although the metamaterials
that have been available so far behave like LHMs
only in the microwave range,
there have been suggestions on how to construct metamaterials
that can operate at optical frequencies, by reducing the sizes
of the inclusions (split rings, chiral or
omega particles) \cite{Panina02} or by using point defects
in photonic crystals as magnetic emitters \cite{Povinelli03}.
A number of potential applications of LHMs have been
proposed, including effective light-emitting devices,
beam guiders, filters, and near-field lenses.
For example, LHMs could be used to realize highly
efficient low reflectance surfaces \cite{Smith00b}
or superlenses which, in principle, can achieve arbitrary subwavelength
resolution \cite{Pendry00}. The intriguing superlense
proposal and the reported observation
of negative refraction \cite{Shelby01sci} have
touched off intensive and enlightening discussions
\cite{discussion,Ziolkowski01,Garcia02}.
More recent experiments \cite{Parazzoli03} seem to confirm the negative
refraction observed in Ref.~\cite{Shelby01sci}.
Nevertheless, there have been still many open
questions about the electrodynamics in
magnetodielectrics, i.e., materials with simultaneously
significant electric and magnetic
properties, including LHMs.

In this paper, we first study the problem of quantization of the
macroscopic electromagnetic field in the presence
of magnetodielectrics, with special emphasis on
LHMs. Apart from the more fundamental interest in the
problem, quantization is required to include nonclassical radiation in the
studies.
Since dispersion and absorption are
related to each other by the Kramers-Kronig
relations, noticeable dispersion implies that
absorption also cannot be omitted in general.
As we will show, quantization of the electromagnetic field
in the presence of dispersing and absorbing magnetodielectrics
can be performed by means of a source-quantity representation
based on the classical Green tensor in a similar way as in
Refs.~\cite{Gruner95,Matloob95,Ho98,Tip97,Stefano00,Knoll01}
for purely dielectric material.

As a simple application of the quantization scheme, we then study
the spontaneous decay of an excited two-level atom
in a dispersing and absorbing magnetodielectric environment,
with special emphasis on an atom in a spherical cavity.
It is well-known that the spontaneous decay of an atom
is influenced by the environment.
If the atom is embedded in a homogeneous, purely electric
medium with real and positive
(frequency-independent) permittivity,
the decay rate without local-field corrections reads
\begin{equation}
\label{intro2}
     \Gamma = n \Gamma_0,
\end{equation}
where $\Gamma_0$ is the decay rate in free space and
$n$ $\!=$ $\!\sqrt{\varepsilon}$ is the refractive index
(see, e.g., \cite{Yablonovitch88,Barnett96} and references therein).
{F}rom energy scaling arguments it can be inferred that
the electric field in a medium corresponds to the electric
field in free space times $1/\sqrt{\varepsilon}$.
{F}rom a mode decomposition one can conclude that
the mode density is proportional to $n^3$. With that,
Eq.~(\ref{intro2}) immediately follows from Fermi's golden rule. Now
if we take into 
account that in the more general case of positive $\varepsilon$ and
$\mu$ the refractive index is
\mbox{$n$ $\!=$ $\!\sqrt{\varepsilon\mu}$}, we conclude that
\begin{equation}
\label{intro3}
     \Gamma = \mu n \Gamma_0.
\end{equation}
Unfortunately, these arguments cannot be used
if, e.g., $\mu$ and $n$ are simultaneously negative.
Basing the calculations on rigorous quantization, we show
that Eq.~(\ref{intro3}) also remains valid in this case.
Moreover, we generalize Eq.~(\ref{intro3}) to the realistic case of 
dispersing and absorbing matter, including local-field effects.

The article is organized as follows.
In Sec.~\ref{Sec:refr}, some general aspects
of the refractive index of a medium whose
permittivity and permeability can
simultaneously become negative are discussed.
Section \ref{quant} is devoted to the quantization
of the electromagnetic field in the presence of a dispersing and
absorbing magnetodielectric medium.
The interaction of the medium-assisted field with additional
charged particles is considered in Sec.~\ref{atoms}
and the minimal-coupling Hamiltonian is given.
In Sec.~\ref{spdecay}, the theory is applied to the
spontaneous decay of an excited two-level atom,
with special emphasis on an atom in a spherical cavity
surrounded by a dispersing and absorbing magnetodielectric.
A summary and some concluding remarks are given
in Sec.~\ref{conclusions}.


\section{Permittivity, permeability, and refractive index}
\label{Sec:refr}

Let us consider a causal linear
magnetodielectric me\-dium characterized by a (relative)
permittivity $\varepsilon({\bf r},\omega)$ and
a (relative) permeability $\mu({\bf r},\omega)$, both of which
are spatially varying, complex functions of frequency
satisfying
the relations
\begin{equation}
\label{e10-1}
\varepsilon({\bf r},-\omega^\ast) = \varepsilon^\ast({\bf r},\omega),
\quad
\mu({\bf r},-\omega^\ast) = \mu^\ast({\bf r},\omega).
\end{equation}
They are holomorphic in the upper complex half-plane without zeros
and approach unity as the frequency goes to infinity,
\begin{equation}
\label{e10-3}
\lim_{|\omega|\to\infty}\varepsilon({\bf r},\omega)
= \lim_{|\omega|\to\infty}\mu({\bf r},\omega) = 1.
\end{equation} 
Since for absorbing media
%
${\rm Im}\,\varepsilon({\bf r},\omega)$ $\!>$ $\!0$,
${\rm Im}\,\mu({\bf r},\omega)$ $\!>$ $\!0$
(see, e.g., Ref.~\cite{Landau}), we may write
\begin{align}
\label{eps1}
        &\varepsilon({\bf r},\omega) = |\varepsilon({\bf r},\omega)|
        e^{i\phi_\varepsilon({\bf r},\omega)},
        &\phi_\varepsilon({\bf r},\omega) \in (0,\pi),
\\
\label{mu1}
	&\mu({\bf r},\omega) = |\mu({\bf r},\omega)|
        e^{i\phi_\mu({\bf r},\omega)}, 
        &\phi_\mu({\bf r},\omega) \in (0,\pi).
\end{align}
The relation $n^2({\bf r},\omega)$ $\!=$
$\!\varepsilon({\bf r},\omega)\mu({\bf r},\omega)$
formally offers two possibilities for 
the (complex) refractive index $n({\bf r},\omega)$,
\begin{equation}
\label{ref1}
        n ({\bf r},\omega)
        = \pm \sqrt{|\varepsilon({\bf r},\omega)\mu({\bf r},\omega)|}
        \,e^{i[\phi_\varepsilon({\bf r},\omega)
        +\phi_\mu({\bf r},\omega)]/2},
\end{equation}
where
\begin{equation}
\label{ref2-3}
0 < [\phi_\varepsilon({\bf r},\omega)+\phi_\mu({\bf r},\omega)]/2 < \pi.
\end{equation}
%
The $\pm$ sign in Eq.~(\ref{ref1}) leads to
${\rm Im}\,n({\bf r},\omega)$ $\!\gtrless$ $\!0$. 
To specify the sign, different arguments can
be used. {F}rom the high-frequency limit of the permittivity and the
permeability, Eq.~(\ref{e10-3}), and the requirement that
%
%
$\lim_{|\omega|\to\infty} n({\bf r},\omega)$ $\!=$ $\!1$
it follows that the $+$ sign is correct \cite{Ziolkowski01}, 
\begin{equation}
\label{ref2-5}
        n ({\bf r},\omega)
        = \sqrt{|\varepsilon({\bf r},\omega)\mu({\bf r},\omega)|}
        \,e^{i[\phi_\varepsilon({\bf r},\omega)
        +\phi_\mu({\bf r},\omega)]/2}.
\end{equation}
The same result can be found from energy arguments \cite{Smith00b}  
[see also the remark following Eq.~(\ref{Gbulk}) in
Sec.~\ref{quant} and Sec.~\ref{nonabsorbing}]. 

{F}rom Eq.~(\ref{ref2-5}) it can be seen that when both
$\varepsilon({\bf r},\omega)$ and $\mu({\bf r},\omega)$ have
negative real parts
[$\phi_\varepsilon({\bf r},\omega),\phi_\mu({\bf r},\omega)\in(\pi/2,\pi)$],
then ${\rm Re}\,n({\bf r},\omega)$ is also negative.
It should be pointed out
that for negative ${\rm Re}\,n({\bf r},\omega)$ it is not
necessary that
${\rm Re}\,\varepsilon({\bf r},\omega)$ and
${\rm Re}\,\mu({\bf r},\omega)$ are simultaneously negative.
For the real part of the refractive index to be negative,
it is sufficient that $[\phi_\varepsilon({\bf r},\omega)
+\phi_\mu({\bf r},\omega)]$ $\!>$ $\!\pi$,
i.e., one of the phases can still be smaller than $\pi/2$,
provided the other one is large enough.
In fact, the definition of LHMs
was originally introduced for
frequency ranges where material absorption is negligible
small, and thus $\varepsilon({\bf r},\omega)$ and $\mu({\bf r},\omega)$
can be regarded as being real \cite{Veselago68}.
In this case
propagating waves can exist provided that both
$\varepsilon({\bf r},\omega)$ and $\mu({\bf r},\omega)$
are simultaneously either positive or negative.
If they
have different signs, then
the refractive index is purely imaginary, and only evanescent waves
are supported. The situation becomes more complicated when
material absorption cannot be disregarded, because there 
is always a nonvanishing real part of the refractive index
(apart from the specific case where $[\phi_\varepsilon({\bf r},\omega)
+\phi_\mu({\bf r},\omega)]$ $\!=$ $\!\pi$).
In the following we refer to a material as being left-handed
if the real part of its refractive index is negative.

\begin{figure}[!t!]
\noindent
\includegraphics[width=1.\linewidth]{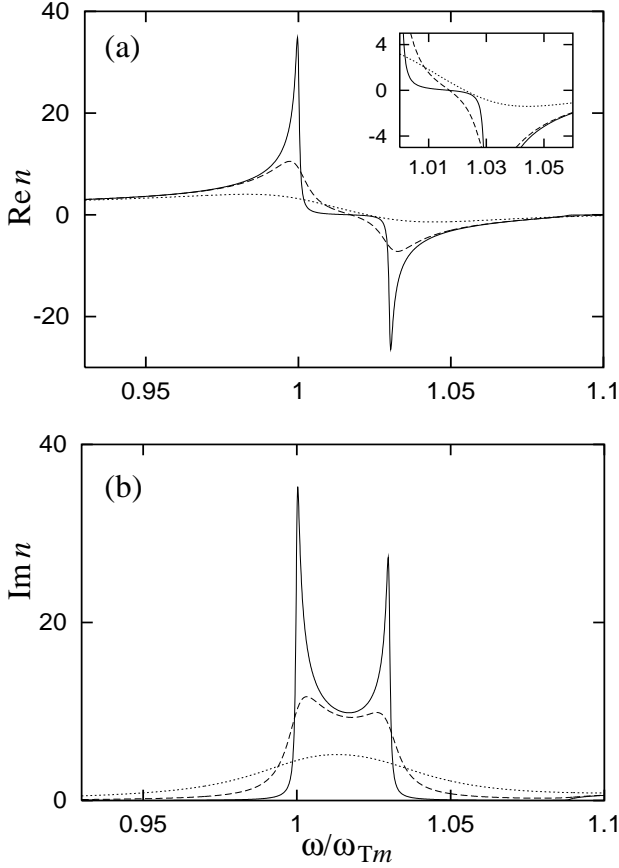}
\caption{%
Real (a) and imaginary (b) parts of the refractive index
$n(\omega)$ as functions of frequency,
with the permittivity $\varepsilon(\omega)$ and the
permeability $\mu(\omega)$ being respectively given
by Eqs.~(\ref{eps}) and (\ref{mu})
[$\omega_{{\rm T}e}$ $\!=$ $\!1.03\,\omega_{{\rm T}m}$;
$\omega_{{\rm P}m}$ $\!=$ $\!0.43\,\omega_{{\rm T}m}$;
$\omega_{{\rm P}e}$ $\!=$ $\!0.75\,\omega_{{\rm T}m}$;
$\gamma_{e}$ $\!=$ $\!\gamma_{m}$
$\!=$ $\!$ $\!0.001\,\omega_{{\rm T}m}$ (solid lines),
$0.01\,\omega_{{\rm T}m}$ (dashed lines), and
$0.05\,\omega_{{\rm T}m}$ (dotted lines)].
The values of the parameters
have been chosen to be similar to those in
Refs.~\cite{Shelby01sci,Garcia02}.
}
\label{refr}
\end{figure}%

In order to illustrate the dependence on frequency of the
refractive index, let us restrict our attention to a
single-resonance permittivity
\begin{equation}
\label{eps}
\varepsilon(\omega) = 1 +
        \frac{\omega_{{\rm P}e}^2}
        {\omega_{{\rm T}e}^2 - \omega^2 - i\omega \gamma_{e}}
\end{equation}
and a single-resonance permeability
\begin{equation}
\label{mu}
\mu(\omega) = 1 +
        \frac{\omega_{{\rm P}m}^2}
        {\omega_{{\rm T}m}^2 - \omega^2 - i\omega \gamma_{m}}\,,
\end{equation}
where $\omega_{{\rm P}e}$, $\omega_{{\rm P}m}$
are the coupling strengths,
$\omega_{{\rm T}e}$, $\omega_{{\rm T}m}$
are the transverse resonance frequencies,
and $\gamma_{e}$, $\gamma_{m}$ are the absorption parameters.
For notational convenience, we have omitted the spatial argument.
Both the permittivity and the permeability satisfy the
Kramers-Kronig relations. Equation (\ref{eps})
corresponds to the well-known (single-resonance) Drude-Lorentz model of
the permittivity. The permeability given by Eq.~(\ref{mu}), which
is of the same type as Eq.~(\ref{eps}), can be derived
by using a damped-harmonic oscillator model for the
magnetization \cite{Ruppin02}. It also occurs in the
magnetic metamaterials constructed recently
\cite{Pendry99,Smith00a,Shelby01sci,Parazzoli03}.

For very small material absorption
($\gamma_{e/m}$ $\!\ll$ $\!\omega_{{\rm P}e/m}$, $\!\omega_{{\rm T}e/m}$),
the permittivity (\ref{eps}) and the permeability (\ref{mu}),
respectively, feature band gaps between the transverse frequency
$\omega_{{\rm T}e}$ and the longitudinal frequency
$\omega_{{\rm L}e}$ $\!=$ $\!\sqrt{\omega_{{\rm T}e}^2
+\omega_{{\rm P}e}^2}$, where \mbox{${\rm Re}\,\varepsilon(\omega)$
$\!<$ $\!0$},
and the transverse frequency 
$\omega_{{\rm T}m}$ and the longitudinal frequency
$\omega_{{\rm L}m}$ $\!=$ $\!\sqrt{\omega_{{\rm T}m}^2
+\omega_{{\rm P}m}^2}$, where ${\rm Re}\,\mu(\omega)$
$\!<$ $\!0$.
With increasing values of $\gamma_e$ and $\gamma_m$
the band gaps are shifted to higher frequencies and smoothed out. 
Figure \ref{refr} shows the dependence on frequency of the
refractive index [Eq.~(\ref{ref2-5})] for the
case of overlapping band gaps and various absorption parameters. 
In particular, if ${\rm max}\,(\omega_{{\rm T}e},\omega_{{\rm T}m})$
$\!<$ $\!\omega$ $\!<$ $\!{\rm min}\,(\omega_{{\rm L}e},
\omega_{{\rm L}m})$, then ${\rm Re}\,\varepsilon(\omega)$ $\!<$ $\!0$
and \mbox{${\rm Re}\,\mu(\omega)$ $\!<$ $\!0$}, and thus a negative
real part of the refractive index is observed.
In Fig.~\ref{refr}, this is the case in the frequency interval where
$1.03$ $\!<$ $\!\omega/\omega_{{\rm T}m}$ $\!<1.088$.
For the chosen parameters, a negative real part of the
refractive index can also be realized for frequencies
slightly smaller than $\omega_{{\rm T}e}$, where
${\rm Re}\,\mu(\omega)$ $\!<$ $\!0$ while
${\rm Re}\,\varepsilon(\omega)$ $\!>$ $\!0$,
as is clearly seen from the inset in Fig.~\ref{refr}(a). 
In this region, however, $|{\rm Re}\,n(\omega)|$ is typically small
whereas ${\rm Im}\,n(\omega)$ is large thereby effectively inhibiting
traveling waves. It is worth noting that a negative real part
of the refractive index is typically observed together with strong
dispersion, so that absorption cannot be disregarded in general.  
On the other hand, increasing absorption smooths the
frequency response of the refractive index thereby making
negative values of the refractive index less pronounced.


\section{The quantized medium-assisted electromagnetic field}
\label{quant}

The quantization of the electromagnetic field in
a causal linear magnetodielectric medium
characterized by both
$\varepsilon({\bf r},\omega)$ and $\mu({\bf r},\omega)$
can be performed by generalizing the theory given in
Refs.~\cite{Ho98,Knoll01} for dielectric media. Let
$\underline{\hat{\bf P}}({\bf r},\omega)$ and
$\underline{\hat{\bf M}}({\bf r},\omega)$, respectively,
be the operators of the polarization and the magnetization
in frequency space. The operator-valued Maxwell equations
in frequency space
then read
\begin{eqnarray}
\label{e1}
&& \bm{\nabla}  \underline{\hat{\bf B}}({\bf r},\omega) = 0,
\\[.5ex]
\label{e2}
&& \bm{\nabla} \underline{\hat{\bf D}}({\bf r},\omega) = 0,
\\[.5ex]
\label{e3}
&& \bm{\nabla} \times \underline{\hat{\bf E}}({\bf r},\omega)
= i\omega \underline{\hat{\bf B}}({\bf r},\omega),
\\[.5ex]
\label{e4}
&& \bm{\nabla} \times \underline{\hat{\bf H}}({\bf r},\omega)
= - i\omega \underline{\hat{\bf D}}({\bf r},\omega),
\end{eqnarray}
where
\begin{eqnarray}
\label{e5}
           \hat{\underline{{\bf D}}}({\bf r},\omega)
           &=& \varepsilon_0
	   \hat{\underline{{\bf E}}}({\bf r},\omega)
	   + \hat{\underline{{\bf P}}}({\bf r},\omega),
\\[.5ex]
\label{e6}
	   \hat{\underline{{\bf H}}}({\bf r},\omega)
	   &=& \kappa_0
	   \hat{\underline{{\bf B}}}({\bf r},\omega) 
           -\hat{\underline{{\bf M}}}({\bf r},\omega)
\end{eqnarray}
[$\kappa_0=\mu_0^{-1}$]. Similarly to the electric constitutive
relation,
\begin{equation}
\label{e7}
\hat{\underline{{\bf P}}}({\bf r},\omega)
           = \varepsilon_0
	   [\varepsilon({\bf r},\omega)-1]
	   \hat{\underline{{\bf E}}}({\bf r},\omega)
	   + \hat{\underline{{\bf P}}}_{\rm N}({\bf r},\omega),
\end{equation}
with $\hat{\underline{{\bf P}}}_{\rm N}({\bf r},\omega)$
being the noise polarization associated with the electric
losses due to material absorption,
we introduce the magnetic constitutive relation 
\begin{equation}
\label{e8}
           \hat{\underline{{\bf M}}}({\bf r},\omega)
	   = \kappa_0 [1 -\kappa({\bf r},\omega)]
	   \hat{\underline{{\bf B}}}({\bf r},\omega) +
           \hat{\underline{{\bf M}}}_{\rm N}({\bf r},\omega),
\end{equation}
where
$\kappa({\bf r},\omega)$ $\!=$ $\!\mu^{-1}({\bf r},\omega)$,
and $\hat{\underline{{\bf M}}}_{\rm N}({\bf r},\omega)$
is the noise magnetization unavoidably associated with magnetic losses.
Recall that for absorbing media
\mbox{${\rm Im}\,\mu({\bf r},\omega)$ $\!>$ $\!0$},
and thus
\mbox{${\rm Im}\,\kappa({\bf r},\omega)$ $\!<$ $\!0$}.
Substituting Eqs.~(\ref{e3}) and (\ref{e5})--(\ref{e8}) into Eq.~(\ref{e4}),
we obtain
\begin{equation}
\label{e11}
      \bm{\nabla} \times \kappa({\bf r},\omega)
      \bm{\nabla}\times
      \underline{\hat{\bf E}}({\bf r},\omega)
       - \frac{\omega^2}{c^2}\varepsilon({\bf r},\omega)
       \underline{\hat{\bf E}}({\bf r},\omega)
       =i\omega\mu_0
       \underline{\hat{\bf j}}_{\rm N}({\bf r},\omega),
\end{equation}
%
where
\begin{equation}
\label{e12}
\hat{\underline{\bf j}}_{\rm N}({\bf r},\omega)
= -i\omega\hat{\underline{\bf P}}_{\rm N}({\bf r},\omega)
+\bm{\nabla} \times
\hat{\underline{\bf M}}_{\rm N}({\bf r},\omega) 
\end{equation}
is the noise current. The noise charge density is given by
$\hat{\underline{\rho}}_{\rm N}({\bf r},\omega)$
$\!=$ $\!- \bm{\nabla}\hat{\underline{\bf P}}_{\rm N}({\bf r},\omega)$,
and the continuity equation holds.
The solution of Eq.~(\ref{e11}) can be given by
\begin{equation}
\label{e13}
\hat{\underline{\bf E}}({\bf r},\omega)
      = i \omega \mu_0
      \int {\rm d}^3 r'
      \,{\bf G}({\bf r},{\bf r}',\omega)
      \,\hat{\underline{\bf j}}_{\rm N}({\bf r}',\omega),
\end{equation}
where $\mbox{\boldmath $G$}({\bf r},{\bf r}',\omega)$ is the
(classical) Green tensor satisfying the equation
\begin{equation}
\label{e14}
\left[
   \bm{\nabla}\times \kappa({\bf r},\omega) \bm{\nabla}\times
   \,-\, \frac{\omega^2}{c^2}\,\varepsilon({\bf r},\omega)
   \right]
   \bm{G}({\bf r},{\bf r}',\omega)
   =  \bm{\delta}({\bf r}-{\bf r}')
\end{equation}
together with the boundary condition at infinity.
It is not difficult to prove that the relation
%
$\bm{G}^{\ast}({\bf r},{\bf r}',\omega)$
$=\bm{G}({\bf r},{\bf r}',-\omega^{\ast})$,
%
which is analogous to the relations (\ref{e10-1}),
is valid. Other useful relations are
(see Appendix \ref{App A})
\begin{equation}
\label{Gprop2}
     G_{ij}({\bf r},{\bf r}',\omega)
     =G_{ji}({\bf r}',{\bf r},\omega)
\end{equation}
and
\begin{eqnarray}
\label{Gprop3}
\lefteqn{
    \int {\rm d}^3 s \Bigl\{
    -{\rm Im}\,\kappa({\bf s},\omega)
    \Bigl[\bm{G}({\bf r},{\bf s},\omega) \times
    \!\overleftarrow{{\bm{\nabla}}}_{\bf s}\Bigr]
    \Bigl[
    {\bm{\nabla}}_{\bf s} \!\times
           \bm{G}^\ast({\bf s},{\bf r}',\omega) \Bigr]
}
\nonumber\\&&\hspace{-2ex}
    + \,\frac{\omega^2}{c^2}\, {\rm Im}\,\varepsilon({\bf s},\omega)
      \,\bm{G}({\bf r},{\bf s},\omega)
      \bm{G}^\ast({\bf s},{\bf r}',\omega)
      \Bigr\}
      = {\rm Im}\,\bm{G}({\bf r},{\bf r}',\omega),
\nonumber\\&&
\end{eqnarray}
where
\begin{equation}
 \Bigl[\bm{G}({\bf r},{\bf s},\omega) \times
    \!\overleftarrow{{\bm{\nabla}}}_{\bf s}\Bigr]_{ij}
 =\epsilon_{jkl}\partial_k^sG_{il}({\bf r},{\bf s},\omega).
\end{equation}
In the simplest case of bulk material, Eq.~(\ref{e14})
implies that the Green tensor can simply be obtained
by multiplying the Green tensor for a bulk
dielectric \cite {Abrikosov,Barnett96,Knoll01}
by $\mu(\omega)$ and replacing $\varepsilon(\omega)$
with $\varepsilon(\omega)\mu(\omega)$,
\begin{equation}
\label{Gbulk}
\begin{split}
 G_{ij}({\bf r},{\bf r}',\omega)=&\mu(\omega)
 \left[\partial^r_i\partial^r_j +
 q^2(\omega)\delta_{ij}({\bf r}-{\bf r}')\right]
\\[.5ex]
 \times\,&\frac{e^{i{\rm Re}\,q(\omega)|{\bf r}-{\bf r}'|}}{4\pi
  q^2(\omega)|{\bf r}-{\bf r}'|}
  \,e^{-{\rm Im}\,q(\omega)|{\bf r}-{\bf r}'|}
  \end{split}
\end{equation}
[$q(\omega)$ $\!=$ $\!n(\omega)\omega/c$].
{F}rom the boundary condition for the Green tensor at
$|{\bf r}$ $\!-$ $\!{\bf r}'|\to\infty$, it follows that
\mbox{${\rm Im}\,n(\omega)$ $\!>$ $\!0$}, which is consistent with
Eq.~(\ref{ref2-5}).   

Analogously to the noise polarization that can be
related to a bosonic vector field $\hat{\bf f}_{e}({\bf r},\omega)$
via
\begin{equation}
\label{e18}
      \hat{\underline{\bf P}}_{\rm N}({\bf r},\omega)
      = i
      \sqrt{\frac{\hbar\varepsilon_0}{\pi}
      {\rm Im}\,\varepsilon({\bf r},\omega)}
      \,\hat{\bf f}_{e}({\bf r},\omega),
\end{equation}
the noise magnetization can be related to a
bosonic vector field $\hat{\bf f}_{m}({\bf r},\omega)$ via
\begin{equation} 
\label{e19}
      \hat{\underline{\bf M}}_{\rm N}({\bf r},\omega)
      =
      \sqrt{-\frac{\hbar\kappa_0}{\pi}
      {\rm Im}\,\kappa({\bf r},\omega)}
      \,\hat{\bf f}_{m}({\bf r},\omega),
\end{equation}
with ($\lambda$, $\!\lambda'$ $\!=$ $\!e,m$)
\begin{eqnarray}
\lefteqn{
\label{com1}
\hspace*{-12ex}
\left[ \hat{f}_{\lambda i}({\bf r},\omega),
\hat{f}_{\lambda' j}^\dagger({\bf r}',\omega')\right]
=  \delta_{\lambda\lambda'}\delta_{ij}\delta({\bf r}-{\bf r}')
\delta(\omega-\omega'),} \\[.5ex]
\label{com2}
&& \left[ \hat{f}_{\lambda i}({\bf r},\omega),
\hat{f}_{\lambda' j}({\bf r}',\omega')\right] =0 .
\end{eqnarray}
Substituting in Eq.~(\ref{e12}) for
$\hat{\underline{\bf P}}_{\rm N}({\bf r},\omega)$ and
$\hat{\underline{\bf M}}_{\rm N}({\bf r},\omega)$
the expressions (\ref{e18}) and (\ref{e19}),
respectively, we may express
$\hat{\underline{\bf j}}_{\rm N}({\bf r},\omega)$
in terms of the bosonic fields $\hat{\bf f}_\lambda({\bf r},\omega)$
as follows: 
\begin{eqnarray}
\label{e20}
\lefteqn{
      \hat{\underline{\bf j}}_{\rm N}({\bf r},\omega)
      = \omega
      \sqrt{\frac{\hbar\varepsilon_0}{\pi}
      \,{\rm Im}\,\varepsilon({\bf r},\omega)}
      \,\hat{\bf f}_{e}({\bf r},\omega)
}      
\nonumber\\[.5ex]&&\hspace{2ex}
      + \, \mbox{\boldmath $\nabla$} \times
      \sqrt{-\frac{\hbar\kappa_0}{\pi}
      \,{\rm Im}\,\kappa({\bf r},\omega)}
      \,\hat{\bf f}_{m}({\bf r},\omega).
\end{eqnarray}
Note that in Eqs.~(\ref{e18}) and (\ref{e19}), respectively,
$\underline{\hat{\bf P}}_{\rm N}({\bf r},\omega)$
and $\underline{\hat{\bf M}}_{\rm N}({\bf r},\omega)$
are only determined up to some phase factors which can be chosen
independently of each other. Here we have them chosen   
such that in Eq.~(\ref{e20}) the coefficients of
$\hat{\bf f}_{e}({\bf r},\omega)$ and 
$\hat{\bf f}_{m}({\bf r},\omega)$ are real.

The $\hat{\bf f}_\lambda({\bf r},\omega)$
and $\hat{\bf f}^\dagger_\lambda({\bf r},\omega)$
can be regarded as being the fundamental 
variables of the system composed
of the electromagnetic field and the medium including the
dissipative system, so that the Hamiltonian can be given by
\begin{equation}
\label{hamf}
\hat{H}
= \sum_{\lambda=e,m}
\int{\rm d}^3 r \int_0^\infty {\rm d} \omega\,\hbar\omega\,
\hat{\bf f}_{\lambda}^\dagger({\bf r},\omega)
\hat{\bf f}_{\lambda}({\bf r},\omega).
\end{equation}
In this approach, the medium-assisted electromagnetic-field is fully
expressed in terms of the
$\hat{\bf f}_\lambda({\bf r},\omega)$
and $\hat{\bf f}^\dagger_\lambda({\bf r},\omega)$.
In particular, the electric-field operator
(in the Schr\"{o}dinger picture) reads
\begin{equation}
\label{e26}
         \hat{{\bf E}}({\bf r}) = \int_0^{\infty} {\rm d}\omega \,
         \hat{\underline{{\bf E}}}({\bf r},\omega) +\mbox{H.c.},
\end{equation}
where $\hat{\underline{{\bf E}}}({\bf r},\omega)$ is given
by Eq.~(\ref{e13}) together with Eq.~(\ref{e20}).
Similarly, the other fields can be
expressed in terms of the
$\hat{\bf f}_\lambda({\bf r},\omega)$
and $\hat{\bf f}^\dagger_\lambda({\bf r},\omega)$, by making
use of Eqs.~(\ref{e3}), (\ref{e7}), (\ref{e8}), (\ref{e18}),
and (\ref{e19}).
It can then be shown (Appendix \ref{App B}) that the
fundamental (equal-time) commutation relations
\begin{eqnarray}
\label{comee}
&\displaystyle
      \bigl[\hat{E}_i({\bf r}),\hat{E}_j({\bf r}')\bigr] = 0 =
      \bigl[\hat{B}_i({\bf r}),\hat{B}_j({\bf r}')\bigr],
\\[.5ex]
\label{comeb}
&\displaystyle
      \bigl[\varepsilon_0\hat{E}_i({\bf r}), \hat{B}_j({\bf r}')\bigr]
      = -i\hbar \epsilon_{ijk} \partial^r_k
      \delta({\bf r}-{\bf r}')
\end{eqnarray}
are preserved.
Furthermore, it can be verified (Appendix \ref{App C})
that in the Heisenberg picture the medium-assisted
electromagnetic-field operators obey the
correct time-dependent Maxwell equations.
%
%

The introduction of a noise magnetization of the type
of Eq.~(\ref{e19}) was first suggested in
Ref.~\cite{Knoll01}, but it was wrongly concluded that
such a noise magnetization and the
noise polarization in Eq.~(\ref{e18}) can be related
to a common bosonic vector field $\hat{\bf f}({\bf r},\omega)$.
Since $\hat{\bf f}_e({\bf r},\omega)$ in Eq.~(\ref{e18})
is an ordinary vector field, whereas $\hat{\bf f}_m({\bf r},\omega)$ in
Eq.~(\ref{e19}) is a pseudo-vector field, the use of a common
vector field would require a relation for the noise
magnetization that is different from Eq.~(\ref{e19})
but must ensure preservation of the commutation
relations (\ref{comee}) and (\ref{comeb}) and lead to
the correct Heisenberg equations of motion.
For the metamaterial considered in
Refs.~\cite{Smith00a,Shelby01sci,Marques02,Grbic02,Parazzoli03},
where the electric properties and the magnetic properties are
provided by physically different material components, the
assumption that the polarization and the magnetization
are related to different basic variables is justified.
It is also in the spirit of Ref.~\cite{Ruppin02}, where the
polarization and the magnetization are caused
by different degrees of freedom.


\section{Interaction of the medium-assisted
field with charged particles}
\label{atoms}

In order to study the interaction of charged particles
with the medium-assisted electromagnetic field, we first
introduce the scalar potential
\begin{equation}
\label{e35-1}
\hat{\varphi}({\bf r})
= \int {\rm d}\omega\, \underline{\hat{\varphi}}({\bf r},\omega)
+ \mbox{H.c.}
\end{equation}
and the vector potential
\begin{equation}
\label{e35-2}
\hat{\bf A}({\bf r})
= \int {\rm d}\omega\, \underline{\hat{\bf A}}({\bf r},\omega)
+ \mbox{H.c.},
\end{equation}
where in the Coulomb gauge, $\underline{\hat{\varphi}}
({\bf r},\omega)$ and
$\underline{\hat{\bf A}}({\bf r},\omega)$
are respectively related to the longitudinal part
$\underline{\hat{\bf E}}{}^\parallel({\bf r},\omega)$
and the transverse part
$\underline{\hat{\bf E}}{}^\perp({\bf r},\omega)$
of $\underline{\hat{\bf E}}({\bf r},\omega)$
[Eq.~(\ref{e13}) together with Eq.~(\ref{e20})]
according to
\begin{eqnarray}
\label{e36}
      &\displaystyle
      -\bm{\nabla} \underline{\hat{\varphi}}({\bf r},\omega) =
       \underline{\hat{\bf E}}{}^\parallel({\bf r},\omega),&
\\[.5ex]
\label{e35}
      &\displaystyle\underline{\hat{\bf A}}({\bf r},\omega) =
      (i\omega)^{-1}
      \underline{\hat{\bf E}}{}^\perp({\bf r},\omega).&
\end{eqnarray}
Similarly, the momentum field $\hat{\bf \Pi}({\bf r})$ that is
canonically conjugated with respect to the vector potential
$\hat{\bf A}({\bf r})$ can be constructed noting that 
$\underline{\hat{\bf\Pi}}({\bf r},\omega)$ $\!=$
$\!-\varepsilon_0\underline{\hat{\bf E}}{^\perp}({\bf r},\omega)$.
Now
the Hamiltonian (\ref{hamf})
can
be supplemented by terms describing the energy of the
charged particles and their interaction energy
with the medium-assisted electromagnetic field
in the same way as in Ref.~\cite{Knoll01}
for dielectric matter. In the minimal-coupling scheme
and for non-relativistic particles,
the total Hamiltonian then reads
\begin{eqnarray}
\label{hamtotal}
\lefteqn{
      \hat{H}= \sum_{\lambda=e,m}
      \int{\rm d}^3 r \int_0^\infty {\rm d}\omega\, \hbar\omega\,
        \hat{\bf f}_{\lambda}^\dagger({\bf r},\omega)
	\hat{\bf f}_{\lambda}({\bf r},\omega)
}
\nonumber\\[.5ex]&&\hspace{2ex}
      +\sum_{\alpha}\frac{1}{2 m_{\alpha}}
      \left[\hat{\bf p}_{\alpha}
      -q_{\alpha}\hat{\bf A}(\hat{\bf r}_{\alpha})\right]^2
\nonumber\\[.5ex]&&\hspace{2ex}
      +\,{\textstyle\frac{1}{2}} \int{\rm d}^3 r \hat{\rho}_{\rm A}({\bf r})
      \hat{\varphi}_{\rm A}({\bf r})
      +\int{\rm d}^3 r \hat{\rho}_{\rm A}({\bf r})
      \hat{\varphi}({\bf r}),
\end{eqnarray}
where $\hat{\bf r}_\alpha$, and $\hat{\bf p}_\alpha$
are respectively the position and the canonical
momentum operator of the $\alpha$th particle
of mass $m_\alpha$ and charge $q_\alpha$.
The first term in Eq.~(\ref{hamtotal})
is the Hamiltonian (\ref{hamf}) of the electromagnetic
field and the medium including the dissipative system.
The second term is the kinetic energy of the
charged particles, and the third term is their
Coulomb energy, with
\begin{eqnarray}
\label{e47}
      &\displaystyle \hat{\rho}_{\rm A}({\bf r})
      =\sum_{\alpha}q_{\alpha}\delta({\bf r}-{\bf r}_\alpha),&
\\[.5ex]
\label{e48}
      &\displaystyle \hat{\varphi}_{\rm A}({\bf r})
      = \int {\rm d}^3 r'\,
      \frac{\hat{\rho}_A({\bf r}')}
      {4\pi\varepsilon_0|{\bf r}-{\bf r}'|}&
\end{eqnarray}
being respectively the charge density and
the scalar potential of the particles. Finally, the last term
is the Coulomb energy of the interaction between the charged
particles and the medium.

Let $\hat{\vec{E}}({\bf r})$ and
$\hat{\vec{B}}({\bf r})$ be respectively the operators of
the electric field and the induction field in the presence
of the charged particles
\begin{eqnarray}
\label{e52}
\hat{\vec{E}}({\bf r}) = 
\hat{{\bf E}}({\bf r}) -
\bm{\nabla} \hat{\varphi}_{\rm A}({\bf r}),
\quad
\hat{\vec{B}}({\bf r})
= \hat{{\bf B}}({\bf r}).
\end{eqnarray}
Accordingly, the displacement field
$\hat{\vec{D}}({\bf r})$ and the magnetic field
$\hat{\vec{H}}({\bf r})$ in the presence of the charged particles
are given by
\begin{eqnarray}
           \hat{\vec{D}}({\bf r})  =
	   \hat{{\bf D}}({\bf r}) -
	   \varepsilon_0 \bm{\nabla} \hat{\varphi}_{\rm A}({\bf r}),
\quad
\label{e51.1}
           \hat{\vec{H}}({\bf r})  =  \hat{{\bf H}}({\bf r}).
\end{eqnarray}
Note that in Eqs.~(\ref{e52}) and (\ref{e51.1})
the electromagnetic
fields
must be thought of as being expressed
in terms of the fundamental fields $\hat{\bf f}_\lambda({\bf r})$
and $\hat{\bf f}^\dagger_\lambda({\bf r})$. {F}rom the
construction of the induction field
and the displacement field
it follows that they obey the time-independent Maxwell equations
\begin{eqnarray}
\label{e54}
      \bm{\nabla} \hat{\vec{B}}({\bf r}) = 0,
\label{e55}
\quad
      \bm{\nabla} \hat{\vec{D}}({\bf r})
      = \hat{\rho}_{\rm A}({\bf r}).
\end{eqnarray}
Further, it can be shown (Appendix \ref{App C}) that
the Hamiltonian (\ref{hamtotal}) generates the correct
Heisenberg equations of motion, i.e., the time-dependent
Maxwell equations
\begin{eqnarray}
\label{e56}
&\displaystyle
      \bm{\nabla}\times \hat{\vec{E}}({\bf r})
      + \dot{\hat{\vec{B}}}({\bf r}) = 0,
\\
\label{e57}
&\displaystyle
      \bm{\nabla}\times \hat{\vec{H}}({\bf r})
      - \dot{\hat{\vec{D}}}({\bf r}) = \hat{\bf j}_{\rm A}({\bf r}),
\end{eqnarray}
where
\begin{equation}
\label{e58}
           \hat{{\bf j}}_{\rm A}({\bf r}) =
           {\textstyle\frac{1}{2}}\sum_\alpha q_\alpha
           \left[ \dot{\hat{\bf r}}_{\alpha}
	   \delta ({\bf r}-\hat{\bf r}_{\alpha})+
	   \delta ({\bf r}-\hat{\bf r}_{\alpha})
	   \dot{\hat{\bf r}}_{\alpha} \right] ,
\end{equation}
and the Newtonian equations of motion for the charged particles
\begin{equation}
\label{newt1}
           \dot{\hat{{\bf r}}}_{\alpha} =  \frac{1}{m_{\alpha}}
           \left[\hat{{\bf p}}_{\alpha}
	   - q_{\alpha}\hat{{\bf A}}(\hat{{\bf r}}_{\alpha})\right] ,
\end{equation}
\begin{equation}
\label{newt2}
           m_{\alpha} \ddot{\hat{{\bf r}}}_{\alpha} =
           q_{\alpha}\left\{\hat{\vec{E}}({\bf r}_{\alpha})
	   + {\textstyle\frac{1}{2}}
           \left[\dot{\hat{{\bf r}}}_{\alpha}\times
	   \hat{{\vec{B}}}({\bf r}_{\alpha}) -
           \hat{{\vec{B}}}({\bf r}_{\alpha})\times
	   \dot{\hat{{\bf r}}}_{\alpha}\right]\right\} .
\end{equation}


\section{Spontaneous decay of an excited two-level atom}
\label{spdecay}

Let us consider a two-level atom (position ${\bf r}_{A}$,
transition frequency $\omega_{A}$)
that resonantly interacts
with the electromagnetic field in the presence of magnetodielectrics
and restrict our attention to the electric-dipole
and the rotating-wave approximations.
By analogy with the case of an atom in the presence of
dielectric material \cite{Ho00,Knoll01}, the Hamiltonian
(\ref{hamtotal}) reduces to
\begin{eqnarray}
\label{e62}
\lefteqn{
        \hat{H} = \sum_{\lambda=e,m}
        \int d^3 r \int_0^\infty {\rm d}\omega \,\hbar\omega\,
	\hat{\bf f}_{\lambda}^\dagger({\bf r},\omega)
	\hat{\bf f}_{\lambda}({\bf r},\omega)
}
\nonumber\\&&
        + \,\hbar\omega_{\rm A}\hat{\sigma}^\dagger\hat{\sigma}
        -\left[
        \hat{\sigma}^\dagger
        {\bf d}_{\rm A}
        \int_0^\infty {\rm d}\omega\,
        \underline{\hat{{\bf E}}}({\bf r}_{\rm A},\omega)
        + {\rm H.c.}\right]\!,
        \quad
\end{eqnarray}
where $\hat{\sigma}$ $\!=$ $\!|l\rangle\langle u|$ and
$\hat{\sigma}^\dagger$ $\!=$ $\!|u\rangle\langle l|$
are the Pauli operators of the two-level atom. Here,
$|l\rangle$ is the lower state whose energy is set equal to zero
and $|u\rangle$ is the upper state of energy $\hbar\omega_{\rm A}$.
Further,
${\bf d}_{\rm A}$ $\!=$ $\!\langle l|\hat{\bf d}_{\rm A}|u\rangle$
$\!=$ $\!\langle u|\hat{\bf d}_{\rm A}|l\rangle$
is the transition dipole moment. 

To study the spontaneous decay of an initially excited
atom, we may look for the system wave function at time $t$
in the form of [$|{\bf 1}_{\lambda}({\bf r},\omega)\rangle$
$\!\equiv$ $\!{\bf f}_{\lambda}^\dagger({\bf r},\omega) |\{0\}\rangle$]
\begin{eqnarray}
\label{e63}
\lefteqn{\hspace{-3ex}
          |\psi(t)\rangle = C_{u}(t)
          e^{-i\tilde{\omega}_{\rm A}t}
          |\{0\}\rangle |u\rangle
}
\nonumber\\&&\hspace{-6ex}
          + \sum_{\lambda=e,m} \int{\rm d}^3r
          \int_0^\infty\!\!\! {\rm d}\omega\,
          e^{-i \omega t}
          {\bf C}_{\lambda l}({\bf r},\omega,t)
	  |{\bf 1}_{\lambda}({\bf r},\omega) \rangle
          |l\rangle,
\end{eqnarray}
where $C_{u}(t)$ and ${\bf C}_{\lambda l}(t)$ are slowly varying
amplitudes and, in anticipation of the environment-induced
transition frequency shift $\delta\omega$
\cite{Ho02},
$\tilde{\omega}_{\rm A} = \omega_{\rm A} - \delta\omega$
is the shifted transition frequency.
The Schr\"{o}dinger equation
$i\hbar\partial_t |\psi(t)\rangle$
$\!=$ $\!\hat{H}|\psi(t)\rangle$
then leads to the set of differential equations
\begin{eqnarray}
\label{e64}
\lefteqn{
          \dot{C}_u(t) = -i\delta\omega C_u(t)
          -\frac{1}{\sqrt{\pi\varepsilon_0\hbar}}
          \int_0^\infty {\rm d}\omega\, \frac{\omega}{c}\,
          e^{-i(\omega-\tilde{\omega}_{\rm A})t}
}
\nonumber \\[.5ex]&&\hspace{-1ex}\times\,
          \int {\rm d}^3 r\,
	  {\bf d}_{\rm A}\biggl\{\frac{\omega}{c}
	  \sqrt{{\rm Im}\,\varepsilon({\bf r},\omega)}
          \,\bm{G}({\bf r}_{\rm A},{\bf r},\omega)
          {\bf C}_{el}({\bf r},\omega,t)
\nonumber\\[.5ex]&&\hspace{-1ex}
	  +\,\sqrt{-{\rm Im}\,\kappa({\bf r},\omega)}
          \left[\bm{G}({\bf r}_{\rm A},{\bf r},\omega) \times
	  \overleftarrow{{\bm{\nabla}}}_{\bf r}\right]
          {\bf C}_{ml}({\bf r},\omega,t)\biggr\},
          \qquad
\end{eqnarray}
\begin{eqnarray}
\label{e65}
\lefteqn{
          \dot{\bf C}_{el}({\bf r},\omega,t) =
          \frac{1}{\sqrt{\pi\varepsilon_0\hbar}}\,
          \frac{\omega^2}{c^2}\,
          \sqrt{{\rm Im}\,\varepsilon({\bf r},\omega)}
          \,e^{i(\omega-\tilde{\omega}_{\rm A})t}
}
\nonumber \\[.5ex]&&\hspace{12ex}
\times\,
          {\bf d}_{\rm A}\bm{G}^\ast({\bf r}_{\rm A},{\bf r},\omega)\,
          C_{u}(t),
\\[.5ex]
\label{e66}
\lefteqn{
          \dot{\bf C}_{ml}({\bf r},\omega,t) =
          \frac{1}{\sqrt{\pi\varepsilon_0\hbar}}\,
          \frac{\omega}{c}\,\sqrt{-{\rm Im}\,\kappa({\bf r},\omega)}
          \,e^{i(\omega-\tilde{\omega}_{\rm A})t}
}
\nonumber \\[.5ex]&&\hspace{12ex}
\times\,
          {\bf d}_{\rm A}\left[
          \bm{G}^\ast({\bf r}_{\rm A},{\bf r},\omega)\times
	  \overleftarrow{{\bm{\nabla}}}_{\bf r}\right]
          C_{u}(t),
          \qquad
\end{eqnarray}
which has to be solved under the initial conditions
%
$C_{u}(0)$ $\!=$ $\!1$, 
${\bf C}_{\lambda l}({\bf r},\omega,0)$ $\!=$ $\!0$.
Formal integrations of Eqs.~(\ref{e65}) and (\ref{e66})
and substitution 
into Eq.~(\ref{e64})
leads to, upon using the 
relation (\ref{Gprop3}),
\begin{equation}
\label{e67}
        \dot{C}_u(t) =-i\delta\omega C_u(t)
	+\int_0^t {\rm d}t'\, K(t-t') C_{u}(t'),
\end{equation}
where
\begin{eqnarray}
\label{e68}
\lefteqn{
        K(t-t') = - \frac{1}{\hbar\pi\varepsilon_0}
        \int_0^\infty \!\!{\rm d}\omega \,\frac{\omega^2}{c^2}
}
\nonumber\\[.5ex]&&\times\,
        e^{-i(\omega-\tilde{\omega}_{\rm A})(t-t')}
	{\bf d}_{\rm A}
	{\rm Im}\,\bm{G}({\bf r}_{\rm A},{\bf r}_{\rm A},\omega)
        {\bf d}_{\rm A}.
\end{eqnarray}
It should be noted that, by integrating with respect to $t$,
the integro-differential equation (\ref{e67}) can
equivalently be expressed in the form of
a Volterra integral equation
of second kind \cite{Ho00}.
Equations (\ref{e67}) and (\ref{e68}) formally look
like those
valid for non-magnetic structures \cite{Ho00}.
Since the matter properties are fully
included in the Green tensor, the results only differ
in the actual Green tensor.

Equations (\ref{e67}) and (\ref{e68}) apply
to an arbitrary coupling regime \cite{Ho00}.
Here, we restrict our attention to the weak-coupling regime, where
the Markov approximation applies.
That is to say, we may replace $C_{u}(t')$ in Eq.~(\ref{e67}) by
$C_{u}(t)$ and approximate the time integral according to
\begin{equation}
\label{e69}
       \int_0^t {\rm d}t'e^{-i(\omega-\tilde{\omega}_{\rm A})(t-t')}
       \to  \zeta(\tilde{\omega}_{\rm A}-\omega)
\end{equation}
[$\zeta(x)$ $\!=$ $\!\pi\delta(x)$ $\!+$ $\!i{\cal P}/x$].
Identifying the principal-part integral with the
transition-frequency shift, we obtain
\begin{equation}
\label{Lambshift}
\delta\omega = \frac{1}{\pi\hbar\varepsilon_0}
   {\cal P}\!\int_0^\infty {\rm d}\omega\,
   \frac{\omega^2}{c^2}\,
   \frac{{\bf d}_{\rm A}
   {\rm Im}\,\bm{G}({\bf r}_{\rm A},{\bf r}_{\rm A},\omega)
   {\bf d}_{\rm A}}
   {\omega-\tilde{\omega}_{\rm A}}\,,
\end{equation}
which, together with Eq.~(\ref{e63}), can be regarded as being
the self-consistent defining equation for the transition-frequency
shift \cite{Ho02}. 
Equation (\ref{e67}) then yields
%
$C_{u}(t)$ $\!=$ $\!\exp\!(-{\textstyle\frac{1}{2}}\Gamma t)$,
%
where the decay rate $\Gamma$ is given by the formula
\begin{equation}
\label{drate}
         \Gamma = \frac{2\tilde{\omega}_{\rm A}^2}
         {\hbar\varepsilon_0c^2}\,
         {\bf d}_{\rm A}
	 {\rm Im}\,\bm{G}({\bf r}_{\rm A},
         {\bf r}_{\rm A},\tilde{\omega}_{\rm A})
        {\bf d}_{\rm A},
\end{equation}
which is obviously valid independently of the (material)
surroundings of the atom.


\subsection{Nonabsorbing bulk material}
\label{nonabsorbing}

Let us first consider the limiting case of non-absorbing bulk
material, i.e., $\varepsilon(\tilde{\omega}_{\rm A} )$
and $\mu(\tilde{\omega}_{\rm A})$
are assumed to be real.
Using the bulk-material Green tensor
(\ref{Gbulk}), it can easily be proved that
\begin{eqnarray}
\label{e72}
{\rm Im}\,\bm{G}({\bf r}_{\rm A},{\bf r}_{\rm A},\tilde{\omega}_{\rm A})
\hspace{-1ex}&=&\hspace{-1ex}
  \frac{\tilde{\omega}_{\rm A}}{6\pi c}\,
  {\rm Re}\left[\mu(\tilde{\omega}_{\rm A})
  n(\tilde{\omega}_{\rm A})\right] \bm{I}.
\end{eqnarray}
Substitution 
into Eq.~(\ref{drate}) yields the decay rate
\begin{equation}
\label{e73}
       \Gamma = {\rm Re}\left[
       \mu(\tilde{\omega}_{\rm A})n(\tilde{\omega}_{\rm A})\right] \Gamma_0,
\end{equation}
where
\begin{equation}
\label{Gamma0}
\Gamma_0=\frac{\tilde{\omega}_{\rm A}^3d_{\rm A}^2}
{3\hbar\pi\varepsilon_0c^3}
\end{equation}
is the free-space decay rate, but taken at the shifted
transition frequency. Eq.~(\ref{e73}) is in agreement
with Eq.~(\ref{intro3}) obtained from simple arguments on
the change of the energy density and the mode density for
positive and frequency-independent $\varepsilon$ and $\mu$.
Clearly, Eq.~(\ref{e73})
is more general in that it also applies to dispersive
magnetodielectrics. In particular, when
$\varepsilon(\tilde{\omega}_{\rm A})$ and $\mu(\tilde{\omega}_{\rm A})$ have
opposite signs, then the refractive index defined according
to Eq.~(\ref{ref2-5}) is purely imaginary, thereby leading
to \mbox{$\Gamma$ $\!=$ $\!0$}. This is because
the electromagnetic field cannot
be excited at $\tilde{\omega}_{\rm A}$,
so that spontaneous emission is completely inhibited.
Note that material absorption always gives rise
to a finite
value of $\Gamma$, which of course can be very small.

{F}rom Eq.~(\ref{e73}) it is clearly seen that
for non-absorbing LHM, i.e.,
$\varepsilon(\tilde{\omega}_{\rm A})$ $\!<$ $\!0$ and
$\mu(\tilde{\omega}_{\rm A})$ $\!<$ $\!0$,
the now real refractive index must also be negative,
in order to arrive at a non-negative value of the decay rate.
This is yet another strong argument for the choice of
the $+$ sign in Eq.~(\ref{ref1}).


\subsection{Atom in a spherical cavity}
\label{multilayer}

For realistic bulk material,
the imaginary part of the Green tensor at equal positions
is singular \cite {Abrikosov,Barnett96,Knoll01}.
Physically, this singularity is fictitious,
because the atom, though surrounded by matter, is
always localized in a small free-space region.
The Green tensor for such an inhomogeneous system reads
\begin{equation}
\label{e74}
        \bm{G}({\bf r},{\bf r}',\omega)
        = \bm{G}^{\rm V}({\bf r},{\bf r}',\omega)
        + \bm{G}^{\rm S}({\bf r},{\bf r}',\omega) ,
\end{equation}
where $\bm{G}^{\rm V}({\bf r},{\bf r}',\omega)$
is the vacuum Green tensor
and $\bm{G}^{\rm S}({\bf r},{\bf r}',\omega)$ is the
scattering part, which
describes the effect of reflections at the surface of discontinuity.
Using Eq.~(\ref{e74}) together with
${\rm Im}\,\bm{G}^{\rm V}({\bf r},{\bf r},\omega)
= (\omega/6\pi c)\bm{I}$ [cf. Eq.~(\ref{e72})],
we can write the decay rate (\ref{drate}) as
\begin{equation}
\label{e74-1}
\Gamma = \Gamma_0 + \frac{2\tilde{\omega}_{\rm A}^2}
         {\hbar\varepsilon_0c^2}\,
         {\bf d}_{\rm A}
	 {\rm Im}\,\bm{G}^{\rm S}({\bf r}_{\rm A},
         {\bf r}_{\rm A},\tilde{\omega}_{\rm A})
        {\bf d}_{\rm A},
\end{equation}
which is again seen to be valid for any type of material.   
 
Within a `classical' theory of spontaneous emission \cite{Chance78},
a formula of the type (\ref{e74-1}) has
been used in Ref.~\cite{Klimov02} to calculate the decay rate
of an atom near a dispersionless and absorptionless
LHM sphere. `Classical' theory means here, that a classically moving
dipole in the presence of macroscopic bodies is considered, with the
value of $\Gamma_0$ being borrowed from quantum mechanics.
As in Ref.~\cite{Klimov02}, the atomic transition
frequency is commonly understood as being that
in free space. {F}rom Eq.~(\ref{e74-1}) it is seen that
the medium-assisted (i.e., shifted)
frequency $\tilde{\omega}_{\rm A}$ must be used
instead of the free-space frequency $\omega_{\rm A}$, since
both values can differ substantially.
         
Let us apply Eq.~(\ref{e74-1})
to an atom in a free-space region surrounded by
a multilayer sphere.
Using the Green tensor given in Ref.~\cite{Li94}, we obtain
\begin{equation}
\label{radial}
\frac{\Gamma^\perp}{\Gamma_0} = 1 + {\textstyle\frac{3}{2}}
\sum_{n=1}^\infty n(n+1)(2n+1)
\left[\frac{j_n(k_{\rm A}r_{\rm A})}{k_{\rm A}r_{\rm A}}\right]^2
{\rm Re}\,C^N_n 
\end{equation}
for a radially oriented dipole moment
(${\bf d}_{\rm A}$ $\!\parallel$ $\!{\bf r}_{\rm A}$) and
\begin{eqnarray}
\label{tangential}
\frac{\Gamma^\parallel}{\Gamma_0} &=& 1
        + {\textstyle\frac{3}{4}}
	\sum_{n=1}^\infty (2n+1)
        \Biggl[ j_n^2(k_{\rm A}r_{\rm A}) \,{\rm Re}\,C^M_n
\nonumber\\[.5ex]&&
        +\,
	\biggl(\frac{[k_{\rm A}r_{\rm A}j_n(k_{\rm A}r_{\rm A})]'}
	{k_{\rm A}r_{\rm A}}\biggr)^2
        {\rm Re}\,C^N_n
        \Biggr]
\end{eqnarray}
for a tangentially oriented dipole moment
\mbox{(${\bf d}_{\rm A}$ $\!\perp$ $\!{\bf r}_{\rm A}$)}
[the prime indicating the derivative with respect to
$k_{\rm A}r_{\rm A}$, 
\mbox{($k_{\rm A}$ $\!=$ $\!\tilde{\omega}_{\rm A}/c$)}].
In Eqs.~(\ref{radial}) and (\ref{tangential}),
$j_n(z)$ and $h_n^{(1)}(z)$ are the spherical Bessel
and Hankel functions of the first kind, respectively. The coefficients
$C^N_n$ and $C^M_n$ have to be determined through recurrence formulas
\cite{Li94}.


\begin{figure}[!t!]
\noindent
\includegraphics[width=0.9\linewidth]{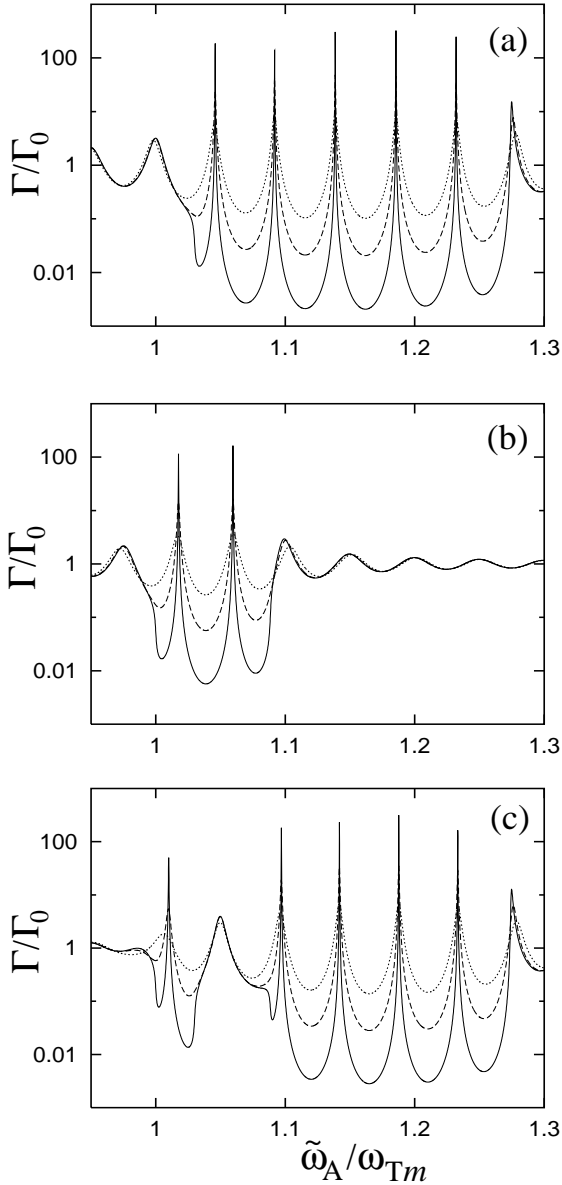}
\caption{
The decay rate $\Gamma$ as a function of
the (shifted) atomic transition frequency
$\tilde{\omega}_{\rm A}$ for an atom at the center of
an empty sphere surrounded by single-resonance matter.
(a) dielectric matter according to Eq.~(\ref{eps})
    [$\omega_{\rm Te}/\omega_{{\rm T}m}$ $\!=$ $\!1.03$;
    $\omega_{\rm Pe}/\omega_{{\rm T}m}$ $\!=$ $\!0.75$; 
    $\gamma_{e}/\omega_{{\rm T}m}$ $\!=$ $\!0.001$ (solid line),
    $0.01$ (dashed line), and
    $0.05$ (dotted line)],
(b) magnetic matter according to Eq.~(\ref{mu})
    [$\omega_{{\rm P}m}/\omega_{{\rm T}m}$ $\!=$ $\!0.43$;
    $\gamma_{m}/\omega_{{\rm T}m}$ $\!=$ $\!0.001$ (solid line),
    $0.01$ (dashed line), and
    $0.05$ (dotted line)], and
(c) magnetodielectric matter according to Eqs.~(\ref{eps})
    and (\ref{mu})
    [the parameters are the same as in (a) and (b)].
    The diameter of the sphere is $2R$ $\!=$ $\!20\,\lambda_{{\rm T}m}$
    ($\lambda_{{\rm T}m}$ $\!=$ $\!2\pi c/\omega_{{\rm T}m}$).
}
\label{R10}
\end{figure}


Equations (\ref{radial}) and (\ref{tangential}) apply to an atom
at an arbitrary position inside a spherical free-space cavity
surrounded by an arbitrary spherical multilayer material
environment. Let us specify the system such that the atom is
situated at the center of the cavity
(i.e., $r_{\rm A}$ $\!=$ $\!0$) and let the surrounding material
homogeneously extend over all the remaining space. For small cavity
radii, the system corresponds to the real-cavity model of local-field
corrections. Making use of the explicit expressions for the
coefficients $C^N_n$ as in Ref.~\cite{Li94} and the fact that for
\mbox{$r_{\rm A}$ $\!=$ $\!0$} only the \mbox{$n$ $\!=$ $1$} term in
Eq.~(\ref{radial}) contributes \cite{Scheel99}, 
we derive from Eq.~(\ref{radial})
\begin{widetext}
\begin{equation}
\label{localf1}
      \frac{\Gamma}{\Gamma_0}
      =
      1 + {\rm Re}\left\{
      \frac{\left[ 1 - i(n+1)z
      - n(n+1)
      {\displaystyle\frac{\mu\!-\!n}{\mu\!-\!n^2}}
      z^2 + i n^2
      {\displaystyle\frac{\mu\!-\!n}{\mu\!-\!n^2}}
      z^3 \right] e^{iz}}
      { -i\sin z - (n\sin z-i\cos z)z
      + \left( \cos z - i
      {\displaystyle\frac{1\!-\!\mu}{\mu\!-\!n^2}}
      n\sin z \right)n z^2
      - (n\sin z+i\mu\cos z )
      {\displaystyle\frac{n^2}{\mu\!-\!n^2}}
      z^3} \right\}
\end{equation}
\end{widetext}
[$\mu$ $\!=$ $\!\mu(\tilde{\omega}_{\rm A})$,
$n$ $\!=$ $\!
n(\tilde{\omega}_{\rm A})$, and
$z$ $\!=$ $\!R\tilde{\omega}_{\rm A}/c$, with $R$ being
the cavity radius]. Obviously, the dipole orientation does not matter
here, and Eqs.~(\ref{radial})
and (\ref{tangential}) lead to exactly the same result.
Equation (\ref{localf1}) is the generalization
of the result derived in Ref.~\cite{Scheel99} for
dielectric matter.

\begin{figure}[!t!]
\noindent
\includegraphics[width=0.9\linewidth]{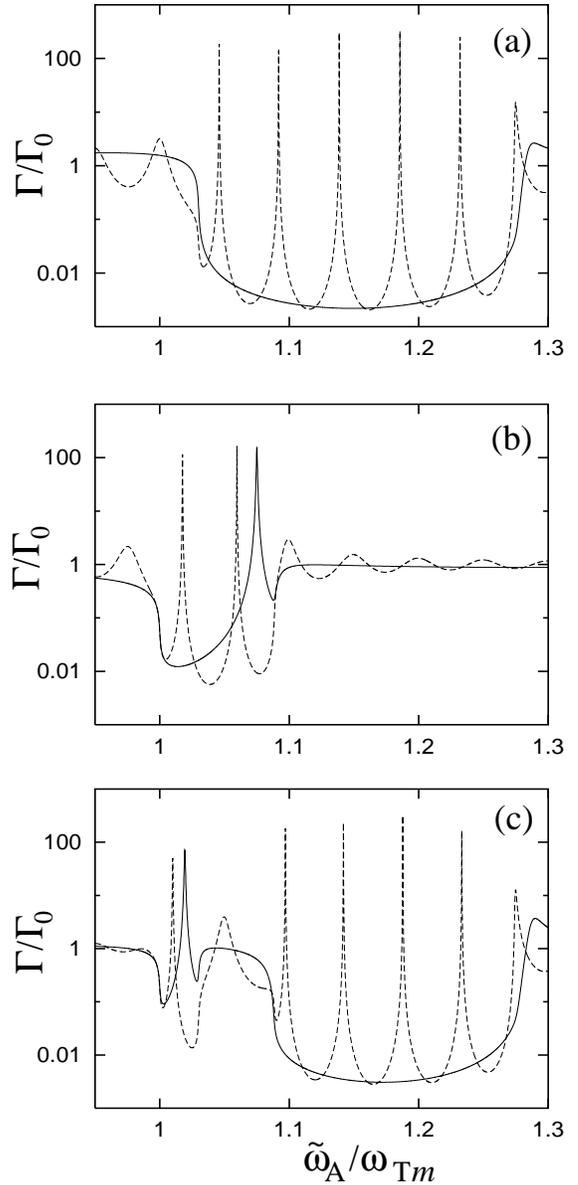}
\caption{
The decay rate $\Gamma$ as a function of
the (shifted) atomic transition frequency
$\tilde{\omega}_{\rm A}$
for an atom at the center of an empty sphere
surrounded by single-resonance matter.
(a) dielectric matter according to Eq.~(\ref{eps}),
(b) magnetic matter according to Eq.~(\ref{mu}), and
(c) magnetodielectric matter according to Eqs.~(\ref{eps})
    and (\ref{mu})
    [$\gamma_{e}/\omega_{{\rm T}m}$ $\!=$
    $\gamma_{m}/\omega_{{\rm T}m}$ $\!=$ $\!0.001$,
    the other parameters are the same as in Fig.~\ref{R10}].
    The values of the sphere diameter are $2R$ $\!=$ $\!20\,\lambda_{{\rm T}m}$
    (dashed lines) and $2R$ $\!=$ $\!1\,\lambda_{{\rm T}m}$
    (solid lines).
}
\label{varR10}
\end{figure}

Figures \ref{R10} -- \ref{R-05} illustrate the
dependence of the decay rate
$\Gamma$ given by Eq.~(\ref{localf1}) 
on the (shifted) transition frequency
for the case of the cavity being surrounded by
(a) purely dielectric matter,
(b) purely magnetic matter, and
(c) magnetodielectric matter
near the band-gap zones.
The permittivity and permeability are given by
Eqs.~(\ref{eps}) and (\ref{mu}), respectively.
In the figures, the dielectric and magnetic band gaps are
assumed to extend from
$\omega_{{\rm T}e}$ $\!=$ $\!1.03\,\omega_{{\rm T}m}$
to \mbox{$\omega_{{\rm L}e}$ $\!\simeq$ $\!1.274\,\omega_{{\rm T}m}$}
and from $\omega_{{\rm T}m}$ to 
$\omega_{{\rm L}m}$ $\!\simeq$ $\!1.088\,\omega_{{\rm T}m}$,
respectively. They overlap in the frequency interval 
$1.03\,\omega_{{\rm T}m}$ $\!<$ $\!\omega$ $\!<$
$\!1.088\,\omega_{{\rm T}m}$.


\subsubsection{Large cavities}
\label{big}

In Fig.~\ref{R10}, a relatively large cavity is considered
($2 R/\lambda_{{\rm T}m}$ $\!=$ $\!20$).
{F}rom Figs.~\ref{R10}(a) and \ref{R10}(b) it is seen that
inside a dielectric or magnetic band gap the decay rate
sensitively depends on the transition frequency. 
Narrow-band enhancement of the spontaneous decay
\mbox{($\Gamma$ $\!>$
$\!\Gamma_0$)} alternates with broadband inhibition
\mbox{($\Gamma$ $\!<$ $\!\Gamma_0$)}. The maxima of enhancement
are observed at the frequencies of the
(propagating-wave) cavity resonances,
the $Q$ factors of which are essentially determined
by the material losses (see the curves for
different values of $\gamma_e$ and $\gamma_m$). 
Note that the cavity resonances as the poles of $\Gamma$
are different for dielectric and magnetic material in general.
{F}rom Fig.~\ref{R10}(c) it is seen that the decay rate of an 
atom surrounded by magnetodielectric matter 
shows a similar behaviour as in Figs.~\ref{R10}(a) and
\ref{R10}(b), provided that the transition frequency is
outside the region of overlapping dielectric and magnetic
band-gap zones. When the transition frequency is in 
the overlapping region of the gaps,
then the medium becomes left-handed. Thus,
a relatively large input-output coupling due
to propagating waves in the medium becomes possible, thereby 
the typical band-gap properties getting lost. As a result, 
neither strong inhibition, nor substantial resonant
enhancement of the spontaneous decay is
observed, as is clearly seen from Fig.~\ref{R10}(c).

In Fig.~\ref{varR10} the results for the cavity in
Fig.~\ref{R10} are compared 
with those observed for a smaller cavity with
$2 R/\lambda_{{\rm T}m}$ $\!=$ $\!1$.
As expected, the number of clear-cut cavity resonances
decreases as the radius of the cavity decreases.
For the smaller of the chosen radii,
just one resonance has survived in the case of the magnetic medium
[Fig.~\ref{varR10}(b)], while the resonances are gone altogether in
the case of the dielectric medium [Fig.~\ref{varR10}(a)].
Accordingly, inhibition of spontaneous decay is typically
observed in the band-gap zones of dielectric and magnetic
matter and in the non-overlapping band-gap region of
magnetodielectric matter.
In contrast, a behaviour quite similar to that in free space
can be observed in the overlapping (left-handed) region.


\subsubsection{Small cavities}
\label{localfield}

\begin{figure}[!t!]
\noindent
\includegraphics[width=0.9\linewidth]{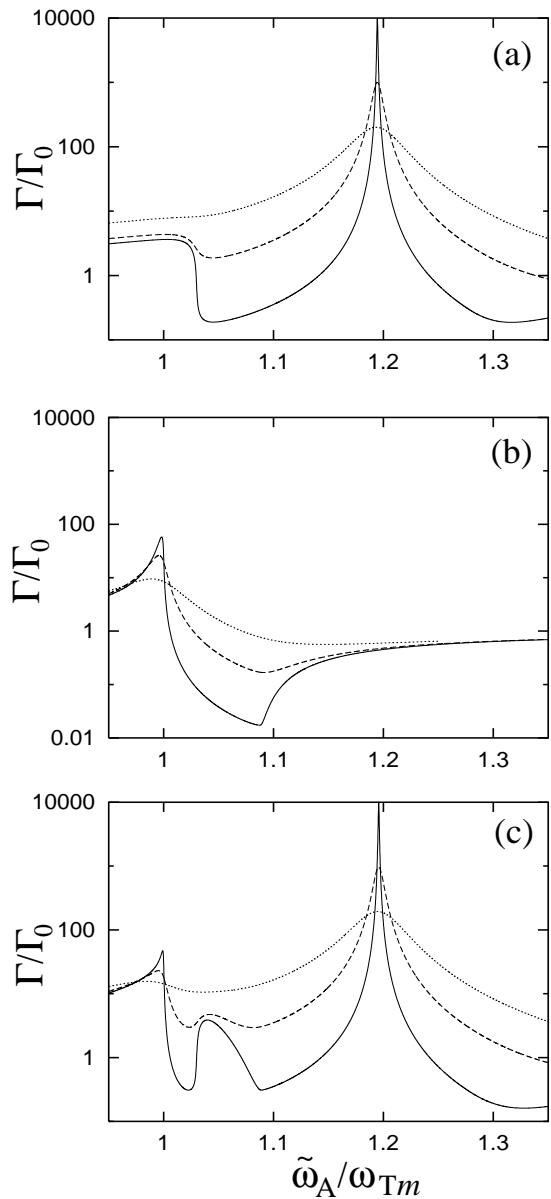}
\caption{%
The decay rate $\Gamma$ as a function of
the (shifted) atomic transition frequency
$\tilde{\omega}_{\rm A}$ for an atom at the center of
an empty sphere surrounded by single-resonance matter.
(a) dielectric matter according to Eq.~(\ref{eps}),
(b) magnetic matter according to Eq.~(\ref{mu}), and
(c) magnetodielectric matter according to Eqs.~(\ref{eps})
    and (\ref{mu}).
    The diameter of the sphere is $2R$ $\!=$ $\!0.1\,\lambda_{{\rm T}m}$.
    The other parameters are the same as in Fig.~\ref{R10}.
}
\label{R-05}
\end{figure}

\begin{figure}[!t!]
\noindent
\includegraphics[width=0.9\linewidth]{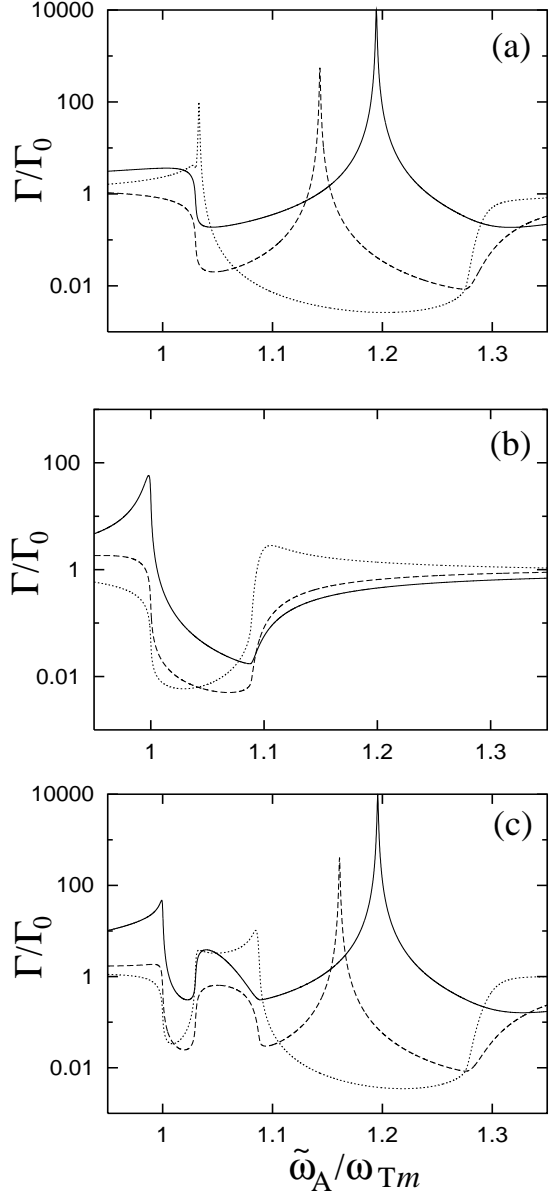}
\caption{
The decay rate $\Gamma$ as a function of
the (shifted) atomic transition frequency
$\tilde{\omega}_{\rm A}$
for an atom at the center of an empty sphere
surrounded by single-resonance matter.
(a) dielectric matter according to Eq.~(\ref{eps}),
(b) magnetic matter according to Eq.~(\ref{mu}), and
(c) magnetodielectric matter according to Eqs.~(\ref{eps})
    and (\ref{mu})
    [$\gamma_{e}/\omega_{{\rm T}m}$ $\!=$
    $\gamma_{m}/\omega_{{\rm T}m}$ $\!=$ $\!0.001$,
    the other parameters are the same as in Fig.~\ref{R10}].
    The values of the sphere diameter are $R$ $\!=$
    $\!0.8\,\lambda_{{\rm T}m}$ (dotted lines),
    $0.4\,\lambda_{{\rm T}m}$ (dashed lines), and
    $0.1\,\lambda_{{\rm T}m}$ (solid lines).
}
\label{varR-05}
\end{figure}

In Fig.~\ref{R-05}, a cavity is considered
whose radius is much smaller than the
transition wavelength ($2 R/\lambda_{{\rm T}m}$ $\!=$ $\!0.1$).
Comparing Fig.~\ref{R-05}(a) with \ref{R-05}(b),
we see that the frequency response of the decay rate in
the dielectric band-gap zone is quite different from that in
the magnetic band-gap zone. In the dielectric band-gap zone
[Fig.~\ref{R-05}(a)], a more or less abrupt decrease of
$\Gamma$ below $\Gamma_0$
with increasing transition frequency is followed by an
increase of $\Gamma$ to a maximum that can substantially
exceed $\Gamma_0$.
In the case of magnetic matter, [Fig.~\ref{R-05}(b)], on
the contrary, only a rather distorted band-gap zone
is observed in which $\Gamma$ monotonously decreases below $\Gamma_0$.
The maximum of enhancement of spontaneous decay in Fig.~\ref{R-05}(a)
is observed at the local-mode resonance associated with
the small cavity, which may be regarded as being a defect of the
otherwise homogeneous dielectric.
This is obviously of the same nature as the donor and acceptor
local modes discussed in Ref. \cite{Yablonovitch91}.
In the regions where the dielectric and magnetic
band-gap zones of the magnetodielectric in
Fig.~\ref{R-05}(c) do not overlap,
the frequency response of the decay rate
is dominated by the respective matter,
i.e., the characteristic features are either dielectric
or magnetic. The situation changes when
the transition frequency is in the overlapping
region, where LHM is realized.
Since this region cannot longer be regarded
as an effectively forbidden zone for propagating
waves, the value of $\Gamma$ can become comparable with
or even bigger than that of $\Gamma_0$.
{F}rom Fig.~\ref{R-05}(c) it is seen that
entering the overlapping region from the magnetic side
stops the decrease of $\Gamma$ on that side, thereby changing
it to an increase. Similarly, the decrease of $\Gamma$ on
the dielectric side stops and changes to an increase when
the overlapping region is entered from the dielectric side.

Figure \ref{varR-05} illustrates the influence
of the cavity radius on the decay rate for small cavities.
Figure \ref{varR-05}(a) reveals that when the value of
$2 R/\lambda_{{\rm T}m}$ changes from
\mbox{$2 R/\lambda_{{\rm T}m}$ $\!=$ $\!0.1$} to
\mbox{$2 R/\lambda_{{\rm T}m}$ $\!=$ $\!0.8$}, then 
the maximum of the spontaneous decay rate associated with the
local-mode resonance in dielectric matter shifts
towards smaller transition frequencies, thereby being reduced.
In the case of magnetic matter,
increasing value of $2 R/\lambda_{{\rm T}m}$ reduces
the distortion of the band-gap zone, as is seen
from Fig.~\ref{varR-05}(b). As expected, the 
frequency response of the decay rate shown in
Fig.~\ref{varR-05}(c) for the case of magnetodielectric
material including LHM combines, in a sense, the respective
curves in Figs.~\ref{varR-05}(a) and \ref{varR-05}(b). 


\subsubsection{Local-field corrections}
\label{lfc}

For an atom in bulk material, the local field
with which the atom really interacts 
can differ from the macroscopic field used in the
derivation of the decay rate of the form given by Eq.~(\ref{e73}).
To include local-field corrections in the rate, one 
can use Eq.~(\ref{localf1}) and let the radius of the cavity
tend to a value which is much smaller than the transition wavelength,
\begin{equation}
\label{e76}
      \frac{R\tilde{\omega}_{\rm A}}{c}
      =\frac{2\pi R}{\lambda_{\rm A}} \ll 1,
\end{equation}
but still much larger than the distances between the medium
constituents to ensure that the macroscopic theory applies. In this
way we arrive at the real-cavity model frequently used in the
literature \cite{Onsager36,Yablonovitch88,Glauber91,Scheel99,Tomas01}.
The results shown in Fig.~\ref{R-05} may be regarded as being
typical of the real-cavity model.   

Expanding $\Gamma$ [Eq.~(\ref{localf1})] in powers of
$z$ $\!=$ $\!R\tilde{\omega}_{\rm A}/c$ we obtain
\begin{eqnarray}
\label{localf2}
\lefteqn{
     \frac{\Gamma}{\Gamma_0} =
     {\rm Re}\,
     \biggl[
     \left(
     \frac{3\varepsilon}{1+2\varepsilon}\right)^2
     \mu n
     \biggr]
     + \frac{9{\rm Im}\,\varepsilon}{|1 + 2\varepsilon|^2}
       \left(\frac{c}{\tilde{\omega}_{\rm A}R}\right)^3
}
\nonumber\\[.5ex]&&
   +\,{\textstyle\frac{9}{5}}\, {\rm Im}\left[
   \frac{\varepsilon(1+3\varepsilon+5\mu\varepsilon)}
   {(1 + 2\varepsilon)^2}\right]
   \left(\frac{c}{\tilde{\omega}_{\rm A}R}\right)
   + O(R),
   \qquad
\end{eqnarray}
which for nonmagnetic media reduces to
results obtained earlier \cite{Scheel99,Tomas01}.
Note that the actual value of $R$,
which is undetermined within the real-cavity model,
should be taken from the experiment. 
Equation (\ref{localf2}) without the
$O(R)$
term has to be employed with great care,
because it fails when, for small absorption, the atomic
transition frequency $\tilde{\omega}_{\rm A}$ becomes close
to a medium resonance frequency such as
$\omega_{{\rm T}e}$ or $\omega_{{\rm T}m}$, thus leading
to a drastic increase of the first term in Eq.~(\ref{localf2}).
The first three terms on the right-hand side
in Eq.~(\ref{localf2}) reproduce
the curves in Fig.~\ref{R-05} sufficiently well, 
except in the vicinities of $\omega_{{\rm T}e}$ and
$\omega_{{\rm T}m}$. In particular, it can easily be checked that
the position of the local-mode-assisted maximum of the
decay rate in the dielectric band-gap zone
is where \mbox{$2\varepsilon(\tilde{\omega}_{\rm A})$
$\!\simeq$ $\!-1$}. 

For transition frequencies that are sufficiently far away from
a medium resonance frequency and,
in case of dielectric and magnetodielectric matter,
the local-mode frequency,
so that material absorption can be disregarded, the first term
in Eq.~(\ref{localf2}) is the leading one, hence
\begin{equation}
\label{localf2-1}
     \Gamma \simeq \left[\frac{3\varepsilon(\tilde{\omega}_{\rm A})}
     {1+2\varepsilon(\tilde{\omega}_{\rm A})}\right]^2
     {\rm Re}\,[\mu(\tilde{\omega}_{\rm A}) n(\tilde{\omega}_{\rm A})]
     \Gamma_0.
\end{equation}
In this case,
the local-field correction
simply results in multiplying the rate obtained for the case
of nonabsorbing bulk material [Eq.~(\ref{e73})] by the factor
$[3\varepsilon/(1$ $\!+$ $\!2\varepsilon)]^2$.
Interestingly, this factor is exactly the same as that for
dielectric material.

Inspection of the second and the third term in Eq.~(\ref{localf2})
shows that such a separation is no longer possible when material
absorption must be taken into account.
It should be pointed out that the second term proportional to
$R^{-3}$ is purely dielectric, whereas the magnetization starts
to come into play only via the third term proportional to $R^{-1}$.
These two terms can be regarded as resulting from
the near-field component and the induction-field component
accompanying the decay of the excited atomic state.
In particular for sufficiently small cavity size and strong
(dielectric) absorption, the second term is the leading one, so that
magnetodielectrics approximately give rise to the same
decay rate as dielectrics: 
\begin{equation}
\label{localf2-2}
\Gamma \simeq \frac{9{\rm Im}\,\varepsilon(\tilde{\omega}_{\rm A})}
{|1 + 2\varepsilon(\tilde{\omega}_{\rm A})|^2}
       \left(\frac{c}{\tilde{\omega}_{\rm A}R}\right)^3
       \Gamma_0. 
\end{equation}
In this case, the decay may be regarded as being purely
radiationless, with the energy being transferred from the
excited atomic state to the surrounding medium mediated
by the near field. 


\section{Summary and conclusions}
\label{conclusions}

It has been shown that the quantization scheme originally
developed for the electromagnetic field in the presence of
dielectric matter described in terms of a spatially varying,
Kramers-Kronig-consistent permittivity
\cite{Gruner95,Matloob95,Ho98,Tip97,Stefano00,Knoll01}
can be extended to causal magnetodielectric
matter, with special emphasis on the recently 
fabricated metamaterials, including LHMs that can
exhibit a negative real part of the refractive index,
thereby leading to a number of unusual properties.
The quantization scheme is based on a source-quantity
representation of the medium-assisted electromagnetic
field in terms of the classical Green tensor and two
independent infinite sets of appropriately chosen bosonic
basis fields of the system that consists of the
electromagnetic field and the medium, including a
dissipative system. We have further shown that the
minimal-coupling Hamiltonian governing the interaction
of the medium-assisted electromagnetic field with additional
charged particles can be obtained from the standard form,
by expressing in it the potentials in terms of the
bosonic basis fields. The theory can serve as basis
for various studies, including generation
and propagation of nonclassical radiation through
magnetodielectric structures,
Casimir forces between magnetodielectric bodies,
or van-der-Waals force between atomic systems and
magnetodielectric bodies.

As an example, we have applied the theory to the problem of
the spontaneous decay of a two-level atom in the presence of
arbitrarily configured, dispersing and absorbing media.
In particular, we have shown
that the theory naturally gives the decay rate and the
frequency shift in terms of the classical Green tensor --
formulas that are valid for any kind of geometry and material.
To be more specific, we have studied the decay rate
of an atom at the center of a cavity surrounded
by a magnetodielectric, assuming a single-resonance permittivity
and a single-resonance permeability of Drude-Lorentz type.
LHM is realized for transition frequencies in the region
where the dielectric and magnetic band-gap zones
overlap, thereby the real parts of the permittivity and
permeability becoming negative. When the transition
frequency enter that region from the
dielectric or magnetic side, then the typical band-gap
properties such as enhancement of the spontaneous decay
at the cavity resonances and inhibition between them
gets lost and a decay rate comparable with that in free
space can be observed. The calculations have been
performed for both large and small cavities.
In particular, if the diameter of the cavity becomes small
compared to the transition wavelength of the atom, the system
reduces to the real-cavity model for including
local-field corrections in the decay rate of the atom
in bulk material. We have discussed this case
in detail both analytically and numerically and made
contact with the results obtained from simple mode-decomposition
arguments in case of positive permittivity and permeability. 

For simplicity, all the calculations have been performed for
isotropic magnetodielectric material, by assuming a scalar
permittivity and a scalar permeability. The extension to
anisotropic material is straightforward. It can be done
in essentially the same way as for anisotropic dielectric
material, by first transforming the permittivity and
permeability tensors into their diagonal forms.
   
\begin{acknowledgments}
We thank Reza Matloob and Adriaan Tip for discussions.
D.G.W. acknowledges discussions with Falk Lederer.
S.Y.B. is grateful for being granted a
Th\"{u}ringer Landesgraduiertenstipendium. S.S. was partly funded by a
Feodor-Lynen-fellowship of the Alexander~von~Humboldt foundation.
This work was supported by the Deutsche Forschungsgemeinschaft and the
EPSRC.
\end{acknowledgments}


\appendix
\section{Some properties of the Green tensor}
\label{App A}

{F}ollowing Ref.~\cite{Knoll01}, we regard the Green tensor as
being the matrix elements in the position basis of a
tensor-valued Green operator $\hat{\bm{G}}$ $\!=$ $\!\hat{\bm{G}}(\omega)$
in an abstract single-particle Hilbert space,
$\bm{G}({\bf r},{\bf r}',\omega)=$
$\langle {\bf r}| \hat{\bm{G}} | {\bf r}' \rangle$,
so that Eq.~(\ref{e14}) can be regarded as the
position-representation of the operator equation
$\hat{\bm{H}}\hat{\bm{G}}=\hat{\bm{I}}$, where
\begin{equation}
\label{A5}
    \hat{\bm{H}}= - \hat{\bf p}\times \kappa(\hat{\bf r},\omega)
    \hat{\bf p}\times
    - \frac{\omega^2}{c^2}\varepsilon(\hat{{\bf r}},\omega) \hat{\bm{I}}.
\end{equation}
Using the relations 
$\langle {\bf r}| \hat{\bf r} | {\bf r}' \rangle$
$\!=$ $\!{\bf r} \delta({\bf r}-{\bf r}')$,
$\langle {\bf r}| \hat{\bf p} | {\bf r}' \rangle$
$=$ $\!-i\bm{\nabla} \delta({\bf r}-{\bf r}')$,
and
$\langle {\bf r}|\hat{\bm{I}}| {\bf r}' \rangle$ $\!=$
$\!\bm{\delta}({\bf r}-{\bf r}')$,
we have
\begin{eqnarray}
\label{A6}
\lefteqn{
    \bm{H}({\bf r},{\bf r}',\omega)\equiv
	\langle {\bf r}| \hat{\bm{H}} | {\bf r}' \rangle
}
\nonumber\\[.5ex]&&
    =\bm{\nabla}\! \times\! \kappa({\bf r},\omega) \bm{\nabla} \times
    \delta({\bf r}\!-\!{\bf r}')
    - \frac{\omega^2}{c^2}\,\varepsilon({\bf r},\omega)
    \bm{\delta}({\bf r}\!-\!{\bf r}'),
    \qquad
\end{eqnarray}
which in Cartesian coordinates reads
\begin{eqnarray}
\label{A7}
\lefteqn{
 H_{ij}({\bf r},{\bf r}',\omega)=
 \biggl\{\partial^r_j\kappa({\bf r},\omega)\partial^r_i
}
\nonumber\\[.5ex]&&
  -\left[\partial^r_l\kappa({\bf r},\omega)\partial^r_l
  +\frac{\omega^2}{c^2}\varepsilon({\bf r},\omega)\right]\delta_{ij}
  \biggr\}\delta({\bf r}-{\bf r}').
  \quad
\end{eqnarray}
Since $\hat{\bm{H}}$ is injective and thus an invertible one-to-one
map between vector functions, we can write $\hat{\bm{G}}$ $\!=$
$\!\hat{\bm{H}}^{-1}$. 
Multiplying this equation by $\hat{\bm{H}}$ from the right, we have
\begin{equation}
\label{A8}
 \hat{\bm{G}}\hat{\bm{H}}=\hat{\bm{I}},
\end{equation}
which in the position basis reads
\begin{equation}
\label{A8b}
 \int{\rm d}^3{s}\,
 \langle {\bf r}|\hat{\bm{G}}|{\bf s}\rangle
 \langle {\bf s}|\hat{\bm{H}}|{\bf r'}\rangle
  = \bm{\delta}({\bf r}-{\bf r}').
\end{equation}
Recalling Eq.~(\ref{A7}), we derive, on integrating by parts
and taking into account that the Green tensor vanishes at
infinity,
\begin{eqnarray}
\label{A9}
\lefteqn{
  \int{\rm d}^3 {s}\, G_{ik}({\bf r},{\bf s},\omega)
  H_{kj}({\bf s},{\bf r}',\omega)
}
\nonumber\\[.5ex] && \hspace{-2ex}
=\biggl\{\partial^{r'}_k\kappa({\bf r}',\omega)\partial^{r'}_j
-\left[\partial^{r'}_l\kappa({\bf r}',\omega)\partial^{r'}_l
+\frac{\omega^2}{c^2}\varepsilon({\bf r}',\omega)\right]
\delta_{kj}\bigg\}
\nonumber\\[.5ex] &&\hspace{5ex}\times\,
G_{ik}({\bf r},{\bf r}',\omega)
 = \delta_{ij}({\bf r}-{\bf r}').
\end{eqnarray}
Interchanging the vector indices
$i$ and $j$ and the spatial arguments ${\bf r}$
and ${\bf r}'$, we obtain
\begin{eqnarray}
\label{A10}
\lefteqn{\hspace{-5ex}
 \Big\{\partial^r_k\kappa({\bf r},\omega)\partial^r_i
  -\big[\partial^r_l\kappa({\bf r},\omega)\partial^r_l
  +\frac{\omega^2}{c^2}\varepsilon({\bf r},\omega)\big]
  \delta_{ki}\Big\}
}
\nonumber\\[.5ex]&&\hspace{10ex}\times
  G_{jk}({\bf r}',{\bf r},\omega)
 = \delta_{ij}({\bf r}-{\bf r}'),
\end{eqnarray}
which, according to Eq.~(\ref{e14}), 
is just the defining equation
for $G_{kj}({\bf r},{\bf r}',\omega)$.
Thus, the reciprocity relation (\ref{Gprop2})
is proved valid.

To prove the integral relation (\ref{Gprop3}), we introduce
operators $\hat{\bm{O}}^\ddagger$ by
$\bigl(\hat{O}^\ddagger\bigr)_{ij}$
$=\bigl(\hat{O}_{ji}\bigr)^\dagger$
$= \hat{O}^\dagger_{ji}$.
{F}rom Eq.~(\ref{A8}) it then follows that
\begin{equation}
\label{A12}
   \hat{\bm{H}}^\ddagger\hat{\bm{G}}^\ddagger=\hat{\bm{I}}.
\end{equation}
Multiplying Eq.~(\ref{A8}) from the right by $\hat{\bm{G}}^\ddagger$
and Eq.~(\ref{A12}) from the left by $\hat{\bm{G}}$ and subtracting
the resulting equations from each other, we obtain
\begin{equation}
\label{A13}
   \hat{\bm{G}}\bigl(\hat{\bm{H}}-\hat{\bm{H}}^\ddagger\bigr)
   \hat{\bm{G}}^\ddagger=\hat{\bm{G}}^\ddagger-\hat{\bm{G}},
\end{equation}
which in the position basis reads
\begin{equation}
\label{A14}
\begin{split}
   \int &\!{\rm d}^3 s \!\int \!{\rm d}^3 s'
   \,G_{im}({\bf r},{\bf s},\omega)
   [H_{mn}({\bf s},{\bf s}',\omega)
\\[.5ex]
    &- H^\ast_{nm}({\bf s}',{\bf s},\omega)]
    G^\ast_{nj}({\bf s}',{\bf r}',\omega)
    =-2i {\rm Im}G_{ij}({\bf r},{\bf r}',\omega).
\end{split}
\end{equation}
Note that
$\langle {\bf r}| \hat{H}_{mn}^\ddagger|{\bf r}'\rangle$ $\!=$
$\!H_{nm}^\ast({\bf r}',{\bf r},\omega)$
and $\langle {\bf r}| \hat{G}_{ij}^\ddagger|{\bf r}'\rangle$ $\!=$
$\!G_{ji}^\ast({\bf r}',{\bf r},\omega)$.
Inserting Eq.~(\ref{A7}) into Eq.~(\ref{A14}), after some
manipulation we derive
\begin{eqnarray}
\label{A15}
\lefteqn{
   \int {\rm d}^3 s \,
   \Bigl\{ {\rm Im}\,\kappa\, ({\bf s},\omega)
   \partial^s_n G_{im}({\bf r},{\bf s},\omega)
}
\nonumber\\[.5ex]&&
   \times\,[\partial^s_m G^\ast_{nj}({\bf s},{\bf r}',\omega)
   - \partial^s_n G^\ast_{mj}({\bf s},{\bf r}',\omega)]
\nonumber\\[.5ex]&&\hspace{2ex}
   +\,\frac{\omega^2}{c^2}\, {\rm Im}\,\varepsilon({\bf s},\omega)
   G_{im}({\bf r},{\bf s},\omega)
   G^\ast_{mj}({\bf s},{\bf r}',\omega)\Bigr\}
   \quad
\nonumber\\[.5ex]&&\hspace{4ex} 
   = {\rm Im}G_{ij}({\bf r},{\bf r}',\omega),
\end{eqnarray}
which is just Eq.~(\ref{Gprop3}) in Cartesian coordinates. 

To examine the asymptotic behaviour of the
Green tensor in the upper half of the complex $\omega$-plane
as $|\omega|\to\infty$ and $|\omega|\to 0$,
we introduce the tensor-valued projectors
\begin{equation}
\label{A16}
   \hat{\bm{I}}^\perp=\hat{\bm{I}} - \hat{\bm{I}}^\parallel,
   \quad
   \hat{\bm{I}}^\parallel=
   \frac{\hat{\bf p}\otimes\hat{\bf p}}{\hat{\bf p}^2}
\end{equation}
[note that $\langle{\bf r}|\hat{\bm{I}}^{\perp(\parallel)}
|{\bf r}'\rangle$ $\!=$ $\!\bm{ \delta}^{\perp(\parallel)}
({\bf r}-{\bf r}')$], and decompose $ \hat{\bm{G}}$ as
\cite{Knoll01},
\begin{eqnarray}
\label{A17}
\lefteqn{
    \hat{\bm{G}} = \hat{\bm{H}}^{-1}
    = \hat{\bm{I}}^\parallel
    (\hat{\bm{I}}^\parallel\hat{\bm{H}}\hat{\bm{I}}^\parallel)^{-1}
    \hat{\bm{I}}^\parallel
}    
\nonumber\\[.5ex]&&\hspace{2ex}
    +\, \bigl[\hat{\bm{I}}^\perp
    -\hat{\bm{I}}^\parallel
    (\hat{\bm{I}}^\parallel\hat{\bm{H}}\hat{\bm{I}}^\parallel)^{-1}
    \hat{\bm{I}}^\parallel\hat{\bm{H}}\hat{\bm{I}}^\perp\bigr]
    \hat{\bm{K}}
\nonumber\\[.5ex]&&\hspace{5ex}\times\,
    \bigl[\hat{\bm{I}}^\perp
    - \hat{\bm{I}}^\perp\hat{\bm{H}}\hat{\bm{I}}^\parallel
    (\hat{\bm{I}}^\parallel\hat{\bm{H}}\hat{\bm{I}}^\parallel)^{-1}
    \hat{\bm{I}}^\parallel\bigr],
\end{eqnarray}
where
\begin{eqnarray}
\label{A18}
   \hat{\bm{K}}=\bigl[\hat{\bm{I}}^\perp\hat{\bm{H}}\hat{\bm{I}}^\perp
   - \hat{\bm{I}}^\perp\hat{\bm{H}}\hat{\bm{I}}^\parallel
    (\hat{\bm{I}}^\parallel\hat{\bm{H}}\hat{\bm{I}}^\parallel)^{-1}
    \hat{\bm{I}}^\parallel\hat{\bm{H}}\hat{\bm{I}}^\perp\bigr]^{-1}.
    \quad
\end{eqnarray}
Recalling that
$\varepsilon({\bf r},\omega),\,\mu({\bf r},\omega)$
$\!\to$ $\!1$ as $|\omega|$ $\!\to$ $\!\infty$,
we easily see that the high-frequency limits of $\hat{\bm{H}}$
and $\hat{\bm{G}}$ are the same as for dielectric material,
thus \cite{Knoll01}
%
%
\begin{eqnarray}
\label{A19b}
    \lim_{|\omega|\to\infty}
    \frac{\omega^2}{c^2}\,{\bm{G}}({\bf r},{\bf r}', \omega)
    = -{\bm \delta}({\bf r}-{\bf r}').
\end{eqnarray}
To find the low-frequency limit of $\hat{\bm{G}}$, 
we note that the second term in Eq.~(\ref{A17}) is
regular. To study the first term,
we distinguish between two cases.\\
{(i)} 
The first term of $\hat{\bm{H}}$ in Eq.~(\ref{A5}) is transverse,
\begin{equation}
 \label{A19c}
 -\hat{\bm I}^\parallel \hat{\bf p}\times
 \kappa(\hat{\bf r},\omega)\hat{\bf p}\times
 = -\hat{\bf p}\times
 \kappa(\hat{\bf r},\omega)\hat{\bf p}\times\hat{\bm I}^\parallel
 =0,
\end{equation}
and therefore does not contribute to
$\hat{\bm{I}}^\parallel\hat{\bm{H}}\hat{\bm{I}}^\parallel$.
It then follows that the same low-frequency behaviour
as in the case of dielectric matter is observed, thus \cite{Knoll01}
\begin{equation}
\label{A20}
    \lim_{|\omega|\to 0}
    \frac{\omega^2}{c^2}\,\hat{\bm{G}}
    = -\hat{\bm{I}}^\parallel
    [\hat{\bm{I}}^\parallel \varepsilon(\hat{\bf r},\omega=0)
     \hat{\bm{I}}^\parallel]^{-1}\hat{\bm{I}}^\parallel,
\end{equation}
i.e., because $\varepsilon({\bf r},\omega=0)$ $\!\neq$ $\!0$, 
\begin{equation}
\label{A21}
 \lim_{|\omega|\to 0}\frac{\omega^2}{c^2}\,
 \bm{G}({\bf r},{\bf r}',\omega) =
 \bm{M}, \qquad M_{ij}<\infty.
\end{equation}
{(ii)} 
Equation (\ref{A19c}) is not valid, so that the
first term of $\hat{\bm{H}}$ in Eq.~(\ref{A5})
contributes to $\hat{\bm{I}}^\parallel\hat{\bm{H}}
\hat{\bm{I}}^\parallel$.
Since \mbox{$\mu({\bf r},\omega=0)$ $\!\neq$ $\!0$} [thus
$\kappa({\bf r},\omega$ $\!=$ $\!0)$ being finite], we find that 
\begin{equation}
\label{A23}
 \lim_{|\omega|\to 0}
 \bm{G}({\bf r},{\bf r}',\omega) = \bm{N},
 \qquad N_{ij}<\infty.
\end{equation}


\section{Commutation relations}
\label{App B}

By using Eqs.~(\ref{e13}), (\ref{e20}),
the commutation relations (\ref{com1}) and (\ref{com2}), and
the integral relation (\ref{Gprop3}), we derive
\begin{eqnarray}
\label{B1}
\lefteqn{
      \bigl[\underline{\hat{E}}_i({\bf r},\omega),
      \underline{\hat{E}}^\dagger_j({\bf r}',\omega')\bigr]
}
\nonumber\\[.5ex]&&\hspace{2ex}      
      = \frac{\hbar\omega^2}{\pi\varepsilon_0 c^2}
      \,{\rm Im}\,G_{ij}({\bf r},{\bf r}',\omega)
      \,\delta(\omega-\omega'),
\end{eqnarray}
\begin{equation}
\label{B1-1}
      \bigl[\underline{\hat{E}}_i({\bf r},\omega),
      \underline{\hat{E}}_j({\bf r}',\omega')\bigr]
      =\bigl[\underline{\hat{E}}_i^\dagger({\bf r},\omega),
      \underline{\hat{E}}^\dagger_j({\bf r}',\omega')\bigr]
      =0.
\end{equation}
{F}rom Eqs.~(\ref{e26}),
(\ref{B1}), and (\ref{B1-1}) it is easily seen that the commutation
relations (\ref{comee}) are valid.
Moreover, we find that,
on recalling
that $\bm{G}^{\ast}({\bf r},{\bf r}',\omega)$
$=\bm{G}({\bf r},{\bf r}',-\omega^{\ast})$,
\begin{alignat}{1}
\label{B3}
      \bigl[\varepsilon_0&\hat{E}_i({\bf r}),\hat{A}_j({\bf r}')\bigr]
\nonumber\\[.5ex]&
      = \int {\rm d}^3 s
      \biggl[ \frac{2i\hbar}{\pi}
      \int_0^\infty {\rm d}\omega\, \frac{\omega}{c^2}\,
      {\rm Im}\,G_{ik}({\bf r},{\bf s},\omega) \biggr]
      \delta_{kj}^\perp ({\bf s}-{\bf r}')
\nonumber\\[.5ex]&
     = \int {\rm d}^3 s \biggl[ \frac{\hbar}{\pi} \, {\cal P}
     \int_{-\infty}^\infty {\rm d}\omega \,\frac{\omega}{c^2}\,
      G_{ik}({\bf r},{\bf s},\omega) \biggr]
      \delta_{kj}^\perp ({\bf s}-{\bf r}')
\nonumber
\\[.5ex]&
     =  \frac{\hbar}{\pi} \,{\cal P}
     \int_{-\infty}^\infty \frac{{\rm d}\omega}{\omega}
     \,\frac{\omega^2}{c^2}\,
      \langle{\bf r}|\hat{\bm G}\hat{\bm I}^\perp|{\bf r}'\rangle_{ij}
\end{alignat}
(${\cal P}$, principal part). Since
the Green tensor is analytic in the upper half of the
complex $\omega$-plane
with the asymptotic behaviour according to Eq.~(\ref{A19b}),
the frequency integral in Eq.~(\ref{B3})
can be evaluated by contour integration along an infinitely
small half-circle around $\omega$ $\!=$ $\!0$, and along an
infinitely large half-circle $|\omega|$ $\!\to$ $\!\infty$.
Taking into account that the Green tensor either has only
longitudinal components in the limit $|\omega|$ $\!\to$ $\!0$, cf.
Eq.~(\ref{A20}), and hence $\omega/c\hat{\bm G}\hat{\bm I}^\perp$
$\!\to$ $\!0$, or is well-behaved, cf. Eq.~(\ref{A23}), we see that
the integral along the infinitely small half-circle vanishes.
Recalling Eq.~(\ref{A19b}), we then readily find
\begin{eqnarray}
\label{B4}
      \bigl[\varepsilon_0\hat{E}_i({\bf r}),\hat{A}_j({\bf r}')\bigr]
      = i\hbar \delta_{ij}^\perp ({\bf r}-{\bf r}').
\end{eqnarray}
Since $\hat{\bf B}({\bf r})$ $\!=$ $\!{\bm \nabla}\times
\hat{\bf A}({\bf r})$, from Eq.~(\ref{B4}) it follows that
\begin{eqnarray}
\label{B4c}
 \bigl[\varepsilon_0 \hat{E}_i({\bf r}),\hat{B}_j({\bf r}')\bigr]
      &\hspace{-1ex}=&\hspace{-1ex}
      -i\hbar\epsilon_{ijk} \partial^r_k \delta({\bf r}-{\bf r}'),
\end{eqnarray}
i.e., Eq.~(\ref{comeb}).
It
is then  not difficult to see that the commutation relation
(\ref{B4}) implies 
\begin{equation}
\label{Bc4-1}
      \bigl[\hat{\varphi}({\bf r}),\hat{A}_i({\bf r}')\bigr] =
      \bigl[\hat{\varphi}({\bf r}),\hat{B}_i({\bf r}')\bigr] = 0.
\end{equation}
%
%

To evaluate commutators involving the displacement field
and the magnetic field, recall Eqs.~(\ref{e5}) -- (\ref{e8}).
%
Using the relations presented above, we derive
\begin{eqnarray}
\label{B7}
&&\bigl[\hat{D}_i({\bf r}),\hat{D}_j({\bf r}')\bigr] = 0,
\\[.5ex]
\label{B8}
&&\bigl[\hat{H}_i({\bf r}),\hat{H}_j({\bf r}')\bigr] = 0,
\end{eqnarray}
and
\begin{equation}
\label{B8b}
\bigl[\hat{D}_i({\bf r}),\mu_0\hat{H}_j({\bf r}')\bigr] =
-i\hbar\epsilon_{ijk}\partial^r_k\delta({\bf r}-{\bf r}').
\end{equation}
Note that
Eq.~(\ref{B8b}) follows by using
similar arguments as in the derivation of Eq.~(\ref{B4}) from
Eq.~(\ref{B3}).
A similar calculation leads to  
\begin{eqnarray}
\label{B9}
     &\displaystyle\bigl[\hat{D}_i({\bf r}),\hat{A}_j({\bf r}')\bigr]
      = i\hbar \delta_{ij}^\perp ({\bf r}-{\bf r}'),&
\\[.5ex]      
\label{B9b}
     &\displaystyle\bigl[\hat{H}_i({\bf r}),\hat{A}_j({\bf r}')\bigr]
     =\bigl[\hat{D}_i({\bf r}),\hat{\varphi}({\bf r}')\bigr]
     =\bigl[\hat{H}_i({\bf r}),\hat{\varphi}({\bf r}')\bigr]
     = 0.&
\nonumber\\&&     
\end{eqnarray}
By combining Eqs.~(\ref{B9}) -- (\ref{B9b})
with Eqs.~(\ref{B4}) and (\ref{Bc4-1}), 
it is not difficult to verify that 
polarization and magnetization commute with the introduced potentials
as well as among themselves.
%
%
%

\section{Heisenberg equations of motion}
\label{App C}

By using the Hamiltonian (\ref{hamtotal}) and recalling
the definitions of the medium-assisted field quantities in
terms of the basic fields $\hat{f}_{\lambda i}({\bf r},\omega)$
the basic-field commutation relations 
(\ref{com1}) and (\ref{com2}), the commutation relations
that have been derived from them, and the standard
commutation relations for the particle coordinates and
canonical momenta, it is straightforward to
prove that the theory yields both the correct Maxwell equations
(\ref{e56}) and (\ref{e57}) and the correct Newtonian equation of
motion (\ref{newt2}).

Let us begin with the Maxwell equations. We derive
on recalling Eqs.~(\ref{e35}) and (\ref{e26}) together
with Eqs.~(\ref{e13}) and (\ref{e20}) and the
commutation relations (\ref{com1}) and (\ref{com2}),
\begin{eqnarray}
\label{C0}
\dot{\hat{\vec{B}}}({\bf r},t)
   &\hspace{-1ex}=&\hspace{-1ex}
   \frac{1}{i\hbar}\,\bigl[\hat{\vec{B}}({\bf r},t),\hat{H}\bigr]
\nonumber\\[.5ex]
   &\hspace{-1ex}=&\hspace{-1ex} \bm{\nabla}\!\times\!\int_0^\infty \!\!{\rm d}\omega\,
   \frac{1}{i\hbar}\bigl[\underline{\hat{\bf A}}({\bf r},\omega),
   \hat{H}\bigr] + {\rm H.c.}
\nonumber\\[.5ex]
   &\hspace{-1ex}=&\hspace{-1ex} -\bm{\nabla}\times \hat{\bf E}({\bf r})
   = -\bm{\nabla}\times \hat{\vec{E}}({\bf r}),
\end{eqnarray}
which is Eq.~(\ref{e56}).
To derive the equation of motion for the displacement field,
we have to consider several commutators according to
\begin{eqnarray}
\label{C1}
\lefteqn{
     \dot{\hat{\vec{D}}}({\bf r})
     = \frac{1}{i\hbar}\,\bigl[\hat{\vec{D}}({\bf r},t),\hat{H}\bigr]
}
\nonumber\\[.5ex] &&
     = \frac{1}{i\hbar} \int \!{\rm d}^3 r'
     \!\int_0^\infty \!\!{\rm d}\omega\, 
     \hbar \omega
      \!\sum_{\lambda=e,m}\bigl[\hat{\bf D}({\bf r}),
      \hat{\bf f}_{\lambda}^\dagger({\bf r}',\omega)
      \hat{\bf f}_{\lambda}({\bf r}',\omega)\bigr]  
\nonumber\\[.5ex] &&\hspace{2ex}
     +\, \frac{1}{i\hbar}\sum_{\alpha}\frac{1}{2 m_{\alpha}}
     \left[\hat{\bf D}({\bf r}), \bigl[\hat{\bf p}_{\alpha}
      -q_{\alpha}\hat{\bf A}(\hat{\bf r}_{\alpha}) \bigr]^2 \right]
\nonumber\\[.5ex] &&\hspace{2ex}
     - \,\frac{\varepsilon_0}{i\hbar}\sum_{\alpha}
     \frac{1}{2 m_{\alpha}}
     \left[\bm{\nabla}\hat{\varphi}_{\rm A}({\bf r}),
     \bigl[\hat{\bf p}_{\alpha}
      -q_{\alpha}\hat{\bf A}(\hat{\bf r}_{\alpha})\bigr]^2\right].
      \qquad
\end{eqnarray}
%
The first commutator in Eq.~(\ref{C1}) can easily be found by
recalling the definitions of displacement and magnetic fields as
\begin{eqnarray}
\label{C1.1}
\lefteqn{
     \frac{1}{i\hbar} \int \!{\rm d}^3 r'
     \!\int_0^\infty \!\!{\rm d}\omega\,
     \hbar \omega
     \!\sum_{\lambda=e,m}\bigl[\hat{\bf D}({\bf r}),
      \hat{\bf f}_{\lambda}^\dagger({\bf r}',\omega)
      \hat{\bf f}_{\lambda}({\bf r}',\omega)\bigr]   
}
\nonumber\\[.5ex]&& 
      = - \int_0^\infty {\rm d}\omega\,i\omega
      \hat{\underline{\bf D}}({\bf r},\omega) + {\rm H.c.}
%
%
%
%
      = \bm{\nabla} \times \hat{\vec{H}}({\bf r}).
      \qquad
\end{eqnarray}
Applying the commutation relation (\ref{B9}),
and recalling the definition of the current density, 
we find that the second term on the right-hand side
of Eq.~(\ref{C1}) can be written as
\begin{equation}
\label{C2}
     \frac{1}{i\hbar}\sum_{\alpha}\frac{1}{2 m_{\alpha}}
     \left[\hat{\bf D}({\bf r}),\bigl[\hat{\bf p}_{\alpha}
      -q_{\alpha}\hat{\bf A}(\hat{\bf r}_{\alpha})\bigr]^2\right]
      = - \hat{\bf j}^\perp_{\rm A}({\bf r}),
\end{equation}
where the Newtonian equation of motion (\ref{newt1})
has been used which
follows directly from the Hamiltonian (\ref{hamtotal}).
Finally, standard commutation relations together with 
the definitions of the scalar potential,
Eqs.~(\ref{e47}) and (\ref{e48}),
and the current density, Eq.~(\ref{e58}) together with Eq.~(\ref{newt1}),
lead to
\begin{equation}
\label{C3}
     - \frac{\varepsilon_0}{i\hbar}\sum_{\alpha}\frac{1}{2 m_{\alpha}}
     [\bm{\nabla}\hat{\varphi}_{\rm A}({\bf r}),[\hat{\bf p}_{\alpha}
      -q_{\alpha}\hat{\bf A}(\hat{\bf r}_{\alpha})]^2]
      = - \hat{\bf j}^\parallel_{\rm A}({\bf r}).
\end{equation}
Inserting Eqs.~(\ref{C1.1}) -- (\ref{C3}) into Eq.~(\ref{C1}),
we arrive at Eq.~(\ref{e57}).

In order to prove
Eq.~(\ref{newt2}), we  consider the equation
\begin{eqnarray}
\label{C4}
\lefteqn{
     m_{\alpha} \ddot{\hat{{\bf r}}}_{\alpha}
     = \frac{1}{i\hbar} \bigl[\hat{\bf p}_{\alpha}
      -q_{\alpha}\hat{\bf A}(\hat{\bf r}_{\alpha}), \hat{H}\bigr]
}
\nonumber\\[.5ex]&&      
     = -\frac{q_{\alpha}}{i\hbar} \!\int\! {\rm d}^3 r
       \!\int_0^\infty \!\!{\rm d}\omega\, \hbar \omega
     \!\sum_{\lambda=e,m} 
     \bigl[\hat{\bf A}(\hat{\bf r}_{\alpha}),
        \hat{\bf f}_{\lambda}^\dagger({\bf r},\omega)
	\hat{\bf f}_{\lambda}({\bf r},\omega)\bigr]
\nonumber\\[.5ex]&&\hspace{2ex}
     +\, \frac{1}{i\hbar}\sum_\beta\frac{1}{2 m_{\beta}}
     \left[\hat{\bf p}_{\alpha}
      -q_{\alpha}\hat{\bf A}(\hat{\bf r}_{\alpha}),
      \bigl[\hat{\bf p}_{\beta}
      -q_{\beta}\hat{\bf A}(\hat{\bf r}_{\beta}) \bigr]^2 \right]
\nonumber\\[.5ex]&&\hspace{2ex}
       + \,\frac{1}{2i\hbar} \int {\rm d}^3 r\,
        \bigl[\hat{\bf p}_{\alpha},
	\hat{\rho}_{\rm A}({\bf r})\hat{\varphi}_{\rm A}({\bf r})\bigr]
\nonumber\\[.5ex]&&\hspace{2ex}
       +\, \frac{1}{i\hbar} \int {\rm d}^3 r\,
        \bigl[\hat{\bf p}_{\alpha},
	\hat{\rho}_{\rm A}({\bf r})\hat{\varphi}({\bf r})\bigr].
\end{eqnarray}
The first term on the right-hand side of Eq.~(\ref{C4})
is again
\begin{eqnarray}
\label{C5}
\lefteqn{
     -\frac{q_{\alpha}}{i\hbar} \!\int\! {\rm d}^3 r
     \!\int_0^\infty \!\!{\rm d}\omega\, \hbar \omega
     \!\sum_{\lambda=e,m} 
     \bigl[\hat{\bf A}(\hat{\bf r}_{\alpha}),
        \hat{\bf f}_{\lambda}^\dagger({\bf r},\omega)
	\hat{\bf f}_{\lambda}({\bf r},\omega)\bigr]
}\qquad
\nonumber\\[.5ex]&&\hspace{10ex}
     = i\omega q_{\alpha}\hat{\bf A}(\hat{\bf r}_\alpha)
     = q_{\alpha}\hat{\bf E}^\perp(\hat{\bf r}_\alpha).
     \qquad
\end{eqnarray}
The second term gives rise to two terms, 
\begin{eqnarray}
\label{C8}
\lefteqn{
     \frac{1}{i\hbar}\sum_\beta\frac{1}{2 m_{\beta}}
     \left[\hat{\bf p}_{\alpha},
      \bigl[\hat{\bf p}_{\beta}
      -q_{\beta}\hat{\bf A}(\hat{\bf r}_{\beta}) \bigr]^2 \right]
}      
\nonumber\\[.5ex]&&
     = {\textstyle\frac{1}{2}} q_\alpha
     \left\{
     \dot{\hat{\bf r}}_{\alpha}\hat{\bf A}(\hat{\bf r}_\alpha)
     \otimes\overleftarrow{\bm{\nabla}}
     + \bm{\nabla}\otimes\hat{\bf A}(\hat{\bf r}_\alpha)
     \dot{\hat{\bf r}}_{\alpha}
     \right\}
     \qquad
\end{eqnarray}
and
\begin{eqnarray}
\label{C9}
\lefteqn{
     -\frac{q_{\alpha}}{i\hbar}\sum_\beta\frac{1}{2 m_{\beta}}
     \left[\hat{\bf A}(\hat{\bf r}_{\alpha}),
      \bigl[\hat{\bf p}_{\beta}
      -q_{\beta}\hat{\bf A}(\hat{\bf r}_{\beta}) \bigr]^2 \right]
}      
\nonumber\\[.5ex]&&
     = - {\textstyle\frac{1}{2}} q_\alpha
     \left\{
     \dot{\hat{\bf r}}_{\alpha}
     \bm{\nabla}\otimes\hat{\bf A}(\hat{\bf r}_\alpha)
     +\hat{\bf A}(\hat{\bf r}_\alpha)\otimes
     \dot{\hat{\bf r}}_{\alpha}\overleftarrow{\bm{\nabla}}
     \right\}
     ,\qquad
\end{eqnarray}
and thus
\begin{eqnarray}
\label{C9b}
\lefteqn{
     \frac{1}{i\hbar}\sum_\beta\frac{1}{2 m_{\beta}}
     \left[\hat{\bf p}_{\alpha}
      -q_{\alpha}\hat{\bf A}(\hat{\bf r}_{\alpha}),
      \bigl[\hat{\bf p}_{\beta}
      -q_{\beta}\hat{\bf A}(\hat{\bf r}_{\beta}) \bigr]^2 \right]
}\qquad
\nonumber\\[.5ex]&&
           ={\textstyle\frac{1}{2}}q_\alpha
           \left[\dot{\hat{{\bf r}}}_{\alpha}\times
	   \hat{\vec{B}}({\bf r}_{\alpha}) -
           \hat{\vec{B}}({\bf r}_{\alpha})\times
	   \dot{\hat{{\bf r}}}_{\alpha}\right].
           \qquad
\end{eqnarray}
%
By means of Eqs.~(\ref{e36}) and (\ref{e47}) 
one can see that the last two terms in Eq.~(\ref{C4}) can be rewritten as
\begin{eqnarray}
\label{C6}
&\displaystyle\frac{1}{2i\hbar} \int {\rm d}^3 r\,
\bigl[\hat{\bf p}_{\alpha},
\hat{\rho}_{\rm A}({\bf r})\hat{\varphi}_{\rm A}({\bf r})\bigr] =
- q_{\alpha}\bm{\nabla} \hat{\varphi}_{\rm A}({\bf r}_\alpha),&
\qquad
\\[.5ex]
\label{C7}
&\displaystyle\frac{1}{i\hbar} \int {\rm d}^3 r\,
\bigl[\hat{\bf p}_{\alpha},
\hat{\rho}_{\rm A}({\bf r})\hat{\varphi}({\bf r})\bigr] =
q_{\alpha}  \hat{\bf E}^\parallel(\hat{\bf r}_\alpha).&
\end{eqnarray}
Inserting  Eqs.~(\ref{C5}), (\ref{C9b}) -- (\ref{C7}) into
Eq.~(\ref{C4}) and making use of Eq.~(\ref{e52}), we just
arrive at Eq.~(\ref{newt2}).


\end{document}